\documentclass[pre,preprint,groupedaddress, showkeys,showpacs]{revtex4}

\usepackage{graphicx}
\usepackage{amsmath}
\usepackage{mathrsfs}
\usepackage{color}

\begin{document}

\title{Spontaneous rectification and absolute negative mobility of inertial Brownian particles induced by Gaussian potentials in steady laminar flows}

\author{Jian-Chun Wu$^{1}$} 
\author{Meng An$^{2,3}$}
\author{Wei-Gang Ma$^{3}$}
\affiliation{$^{1}$School of Physics and Electronic Information, Shangrao Normal University, Shangrao 334001, China\\
$^{2}$College of Mechanical and Electrical Engineering, Shaanxi University of Science and Technology, Xi'an 710021, China\\
$^{3}$Key Laboratory for Thermal Science and Power Engineering of Ministry of Education, Department of Engineering Mechanics, Tsinghua University, Beijing 100084, China}

\date{\today}
\begin{abstract}
  \indent We study the transport of inertial Brownian particles in steady laminar flows in the presence of two-dimensional Gaussian potentials. Through extensive numerical simulations, it is found that the transport is sensitively dependent on the external constant force and the Gaussian potential. Within tailored parameter regimes, the system exhibits a rich variety of transport behaviors. There exists the phenomenon of spontaneous rectification (SR), where the directed transport of particles can occur in the absence of any external driving forces. It is found that SR of the particles can be manipulated by the spatial position of the Gaussian potential. Moreover, when the potential lies at the center of the cellular flow, the system exhibits absolute negative mobility (ANM), i.e., the particles can move in a direction opposite to the constant force. More importantly, the phenomenon of ANM induced by Gaussian potential is robust in a wide range of the system parameters and can be further strengthened with the optimized parameters, which may pave the way to the implementation of related experiments.
\end{abstract}


\maketitle

\section {Introduction}
\indent Over the past several decades, the transport of Brownian particles in random environment has played a central role in statistical physics, aiming at obtaining energy from random fluctuations. Generally, Brownian particles diffusion freely or move in the direction of external force in free space. However, the transport may show counter-intuitive behavior in some nonequilibrium systems\cite{Hanggi1,Reimann1}. In ratchet systems, the directed motion of Brownian particles can occur in the absence of external biases\cite{Denisov1,Denisov2}. In order to study the related transport phenomena, researchers proposed some typical ratchet models corresponding to the type of nonequilibrium driving, they are rocking ratchets\cite{Magnasco,Bartussek}, flashing ratchets\cite{Astumian,Borromeo,Bao}, coupled ratchets \cite{Zheng,SouzaSilva}, entropic ratchets\cite{Reguera,Ai,wu1}. In addition, directed transport can also be realized by using static and dynamic force fields, such as the coupling of conservative and nonconservative forces\cite{Zapata}, the transversal ac force\cite{wu2,KYe}, the superimposed driven lattices\cite{Mukhopadhyay1,Mukhopadhyay2}, and the optical ratchet with a rocking driving\cite{Arzola}.

\indent When the system is perturbed by a static force, the transport may exhibit complicated behavior depending on the surrounding environments. Due to the interaction between Brownian particles and complex environments, the system may show a phenomenon of negative differential mobility (NDM)\cite{Jack,Leitmann,Basu,Benichou1,Benichou2,Baiesi,Cividini,Chatterjee}, i.e., the velocity-force relation showing a nonmonotonic behavior. More surprisingly, under specific conditions, the particles can move in the direction opposite to the direction of the driving force, which is called absolute
negative mobility (ANM). Earlier studies have revealed that ANM can be realized in some specific setups, such as semiconductor superlattices\cite{Keay}, coupled Brownian motors\cite{Reimann2,Cleuren1}, and well-designed geometric channels\cite{Eichhorn1,Eichhorn2,Cleuren2}. Shortly afterwards, scientists found that ANM can also be observed in some structures by choosing proper driving and coupling ways\cite{Eichhorn3,Speer1,Speer2}. By embedding the spatial asymmetry into the particle shape, H\"{a}nggi and co-workers\cite{Hanggi2,Ghosh} realized a classical ANM for elongated particles in two-dimensional (2D) separate channels. Machura and co-workers\cite{Machura} studied the motion of inertial Brownian particles in one-dimensionl (1D) symmetric periodic potentials, and found that ANM can be induced by thermal equilibrium fluctuations under the influence of both a time periodic force and a constant force. Based on this work, a large number of extended studies have realized ANM under different conditions\cite{Nagel,Hennig,Kostur,Du,Mulhern,Spiechowicz,Malgaretti,Dandogbessi,Chen,lapik}.

\indent Recently, the phenomenon of ANM has been observed in steady laminar flows\cite{Sarracino,Cecconi1,Cecconi2,Ai2,Reichhardt1}. Sarracino and co-workers\cite{Sarracino} studied the mobility of an inertial tracer in a two-dimensional incompressible laminar flow and observed the phenomena of negative differential mobility and absolute negative mobility, where the velocity field played a key role in these nonlinear behaviors. By applying a 1D periodic potential, Ai and co-workers\cite{Ai2} found that absolute negative mobility can be drastically enhanced by choosing appropriate phase and height of the potential. However, 1D periodic potential inevitably hinders the motion of Brownian particles in 2D space due to the profile of the potential being invariable in the $y$ direction. An alternative way is to apply 2D potentials to overcome the limitation of 1D potential. In this paper, we introduce 2D Gaussian potentials in the cellular flow and study the transport of inertial Brownian particles. Unlike 1D periodic potential\cite{Ai2}, 2D Gaussian potentials only affects the motion of particles in local space, thus may exhibit some interesting behaviors. Different from the previous studies\cite{Sarracino,Ai2}, our system not only can realize ANM with a wider range of the system parameters, but also exhibits the phenomenon of SR.

\section{Model and methods}
\begin{figure}[htbp]
\begin{center}
\includegraphics[width=15cm]{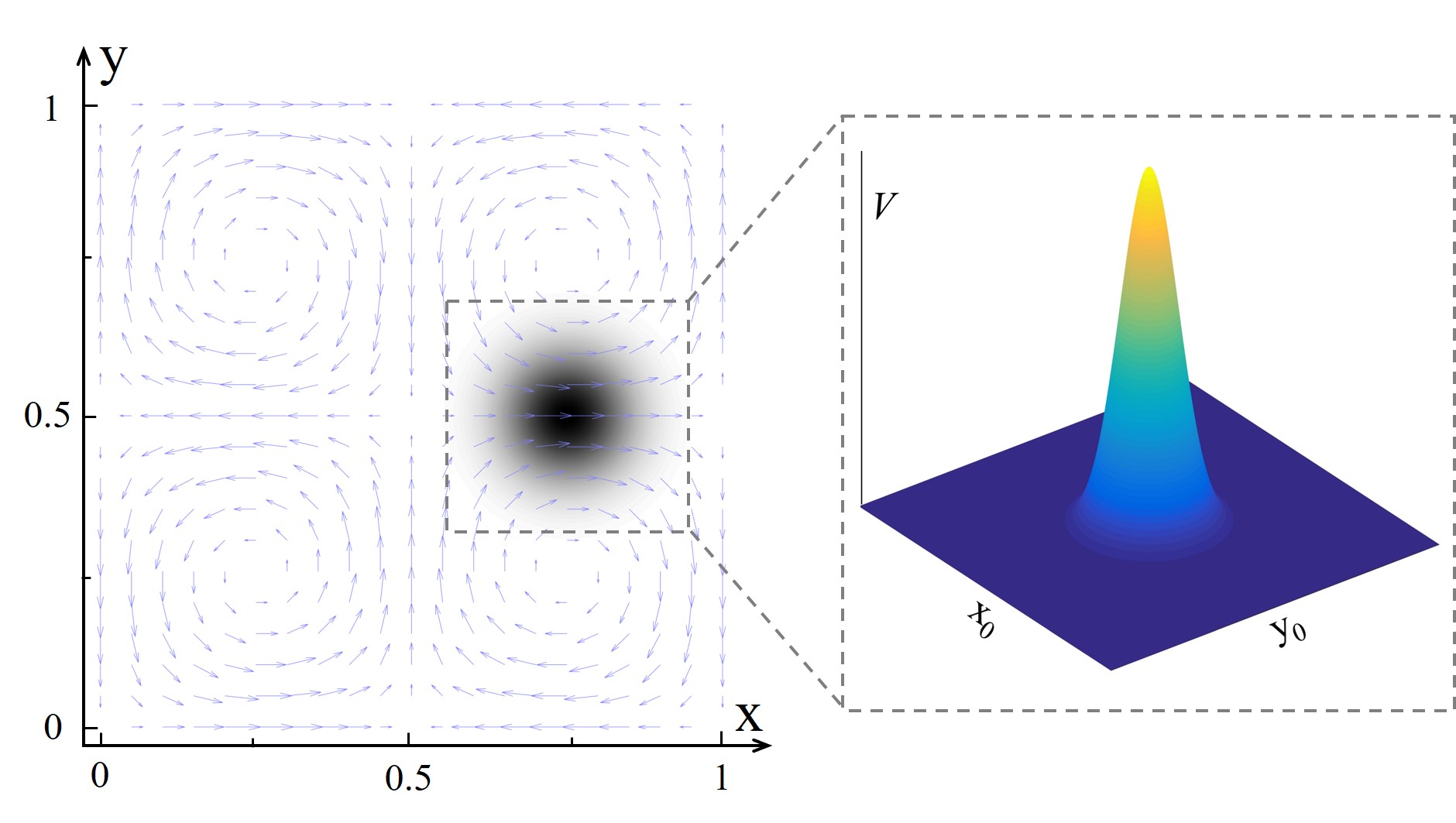}
\caption{Schematic diagrams of the model in one unit cell. The divergenceless cellular flow $(U_{x},U_{y})$ is described by the stream-function $\varphi(x,y)$ in Eq.(5). The Gaussian potential $V(x,y)$ with the spatial coordinates $(x_{0},y_{0})$ in the inset is described by Eq.(6).}\label{1}
\end{center}
\end{figure}

\indent We consider a 2D system of size $L\times L$ with periodic boundary conditions as shown in Fig. 1. In this system, an inertial particle with spatial coordinates $(x,y)$ and velocities $(v_{x},v_{y})$ moves in a divergenceless cellular flow $(U_{x},U_{y})$. The dynamics of the particle can be described by the following equations\cite{Sarracino,Cecconi1,Cecconi2,Ai2},

\begin{equation}
\frac{dx}{dt}=v_{x},
\end{equation}

\begin{equation}
\frac{dy}{dt}=v_{y},
\end{equation}

\begin{equation}
\frac{dv_x}{dt}=-\frac{1}{\tau}(v_{x}-U_{x})+\frac{1}{m}(f+F_{x})+\sqrt{2D}\xi_{x},
\end{equation}

\begin{equation}
\frac{dv_y}{dt}=-\frac{1}{\tau}(v_{y}-U_{y})+\frac{1}{m}F_{y}+\sqrt{2D}\xi_{y}.
\end{equation}

\noindent\\
Here $\tau$ is the Stokes time, $D$ denotes the diffusion coefficient. $\xi_{x}$ and $\xi_{y}$ model white Gaussian noises with zero mean and obey $\langle \xi_{i}(t) \xi_{j}(t')\rangle=\delta_{ij}\delta(t-t')$ for $i,j=x,y$. $m$ is the mass of the inertial Brownian particle and is set to $1$ throughout the work. The divergenceless cellular flow $(U_{x},U_{y})=(\partial \varphi/\partial y, -\partial \varphi/\partial y)$\cite{Tabeling} and the stream-function $\varphi(x,y)$,

\begin{equation}
\varphi(x,y)=\frac{LU_{0}}{2\pi}\sin(\frac{2\pi x}{L}) \sin(\frac{2\pi y}{L}).
\end{equation}

In this paper, the particle is subjected to the force field consists of two parts, a external constant force $f$ along the $x$ direction, and the substrate forces $F_{x}$ and $F_{y}$ from the Gaussian potentials,

\begin{equation}
V(x,y)=A e^{-[(x-x_{0})^2+(y-y_{0})^2]/\varepsilon^2},
\end{equation}
where $A$ and $\varepsilon$ denote the strength and width of the potential, respectively. $x_{0}$ and $y_{0}$ are the spatial coordinates of the potential in one unit cell, the inset of Fig. 1.

\indent In the following, we introduce the characteristic length $L$ and time $L/U_{0}$ by setting $L=1$ and $U_{0}=1$, respectively. In numerical simulations, we integrate Eqs. (1-4) by using the second-order stochastic Runge-Kutta algorithm. Initially, the particles are randomly distributed in the laminar flows. The time step and the total integration time are chosen as $10^{-4}$ and $10^{4}$, respectively, and the data in the paper have been obtained by ensemble averaging over a minimum of $10^{4}$ trajectories.

For simplicity, we only investigate the transport of Brownian particles for the case of $y_{0}=0.5$. Different from the studies of the velocity-force relation\cite{Reichhardt2,Reichhardt3}, the average velocity in the $y$ direction is zero due to the system being completely symmetric along the $y$ direction. Thus, we calculate the average velocity and effective diffusion coefficient along the $x$ direction are, respectively, defined as
            \begin{equation}\label{V}
            \langle v_{x}\rangle=\lim_{t\rightarrow\infty}\frac{\langle x(t)\rangle}{t},
            \end{equation}

            \begin{equation}\label{V}
            D_{x}=\lim_{t\rightarrow\infty}\frac{\langle x^{2}(t)\rangle-\langle x(t)\rangle^{2}}{2t},
            \end{equation}
where $\langle...\rangle$ denotes the average over trajectories of the particle with random initial conditions and noise realizations. The mobility along the $x$ direction is defined as $\mu=\langle v_{x}\rangle/f$.

\section{Results and Discussion}

\begin{figure}[htbp]
\begin{center}
\includegraphics[width=8cm]{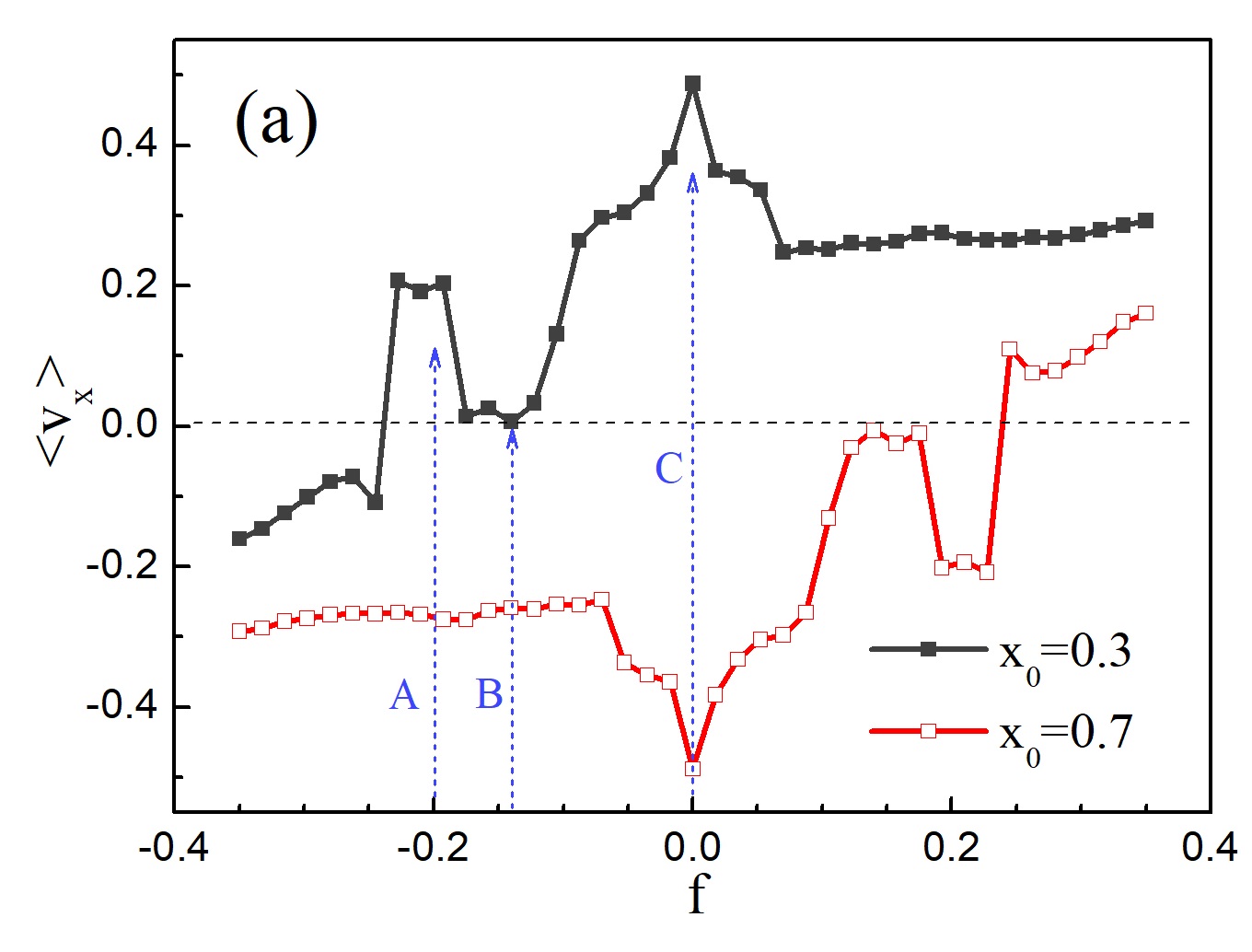}
\includegraphics[width=8cm]{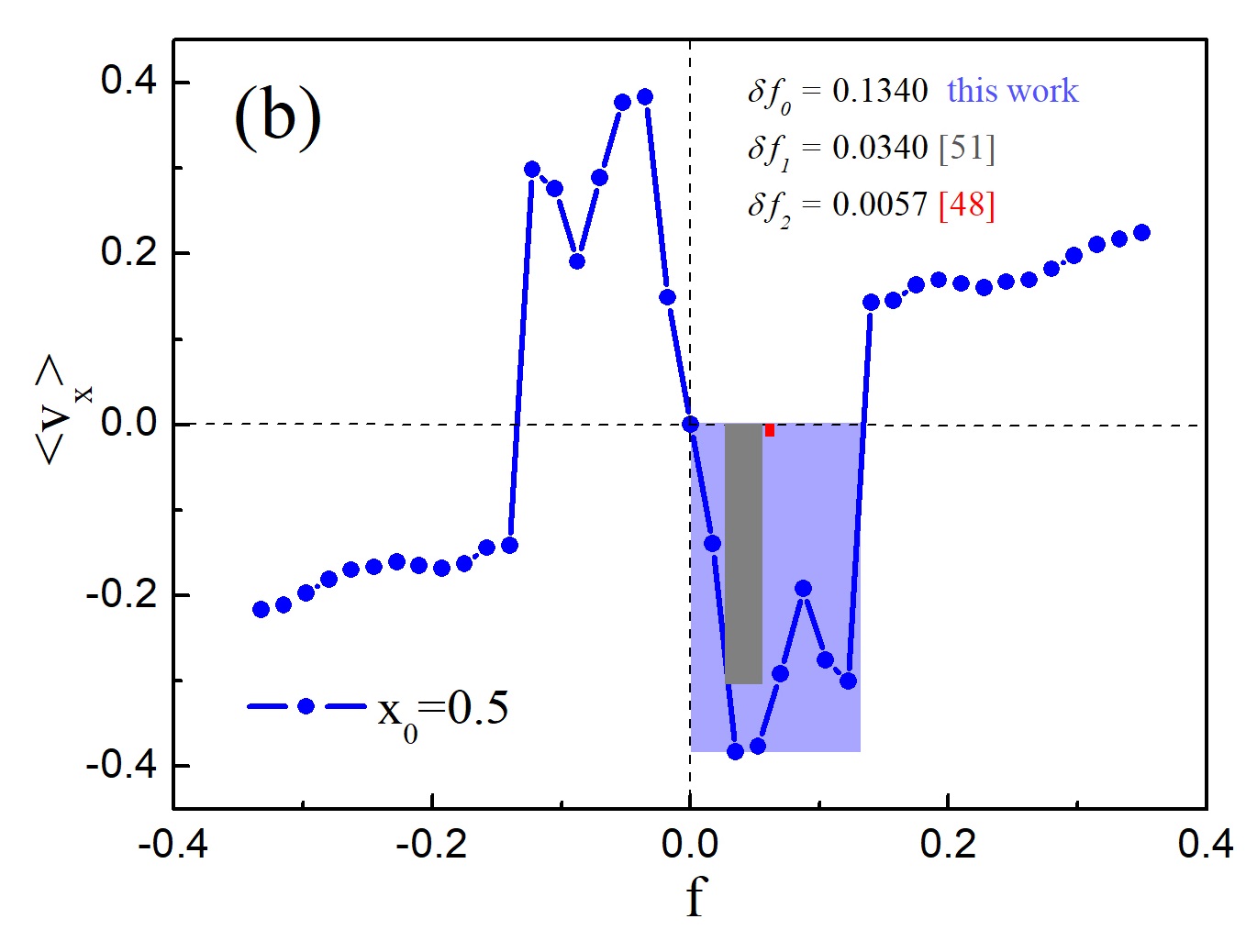}
\caption{
Average velocity $\langle v_{x}\rangle$ as a function of the constant force $f$ for different values of $x_{0}$. The parameters are chosen as $\varepsilon=0.1$, $A=1.0$, $D=10^{-5}$ and $\tau=1.0$.
}\label{1}
\end{center}
\end{figure}

\indent In this section, we mainly focus on how the spatial position of Gaussian potential affects the transport behavior of inertial Brownian particles. The velocity-force curves for different values of $x_{0}$ are plotted in Fig. 2. By introducing Gaussian potentials in the cellular flow, one can observe some peculiar nonlinear behaviors for appropriate $f$. In the case of $x_{0}=0.3$, the transport shows very complex behaviors under the combined action of the Gaussian potential and the driving force. When $f<-0.24$ and $f>0$, the particles move in the direction of the constant force. When $-0.24<f<-0.18$ and $-0.14<f<0$, the direction of transport is opposite to the constant force. More interestingly, we observe the spontaneous rectification of Brownian particles at $f=0$, see the blue arrow $C$ in Fig. 2(a). Usually, the ratchet setup requires three critical ingredients\cite{Reimann1,Denisov1,Denisov2}: spatial or temporal asymmetry, nonlinearity, and nonequilibrium. In the present system, the Gaussian potential can provide the spatial asymmetry. The velocity field can produce the system nonlinearity and break thermodynamical equilibrium. Therefore, directed transport of Brownian particles can be realized at $f=0$. In the case of $x_{0}=0.7$, the transport behavior is completely opposite to that in the case $x_{0}=0.3$ due to the parity symmetry along the $x$ direction.

\indent In addition, we present the velocity-force curves when Gaussian potential locates at the center of the cellular flow ($x_{0}=0.5$) in Fig. 2(b). There is a range of $f$ for which the direction of the transport is opposite to the direction of the constant force, i.e. ANM. By comparison, ANM is greatly enhanced and the regime of $f$ ($0<f<0.134$) for the appearance of ANM is much larger than that in previous studies, such as $0.063<f<0.0687$ (red area)\cite{Sarracino} and $0.026<f<0.06$ (gray area)\cite{Ai2}.

\begin{figure}[htbp]
\begin{center}
\includegraphics[width=6cm]{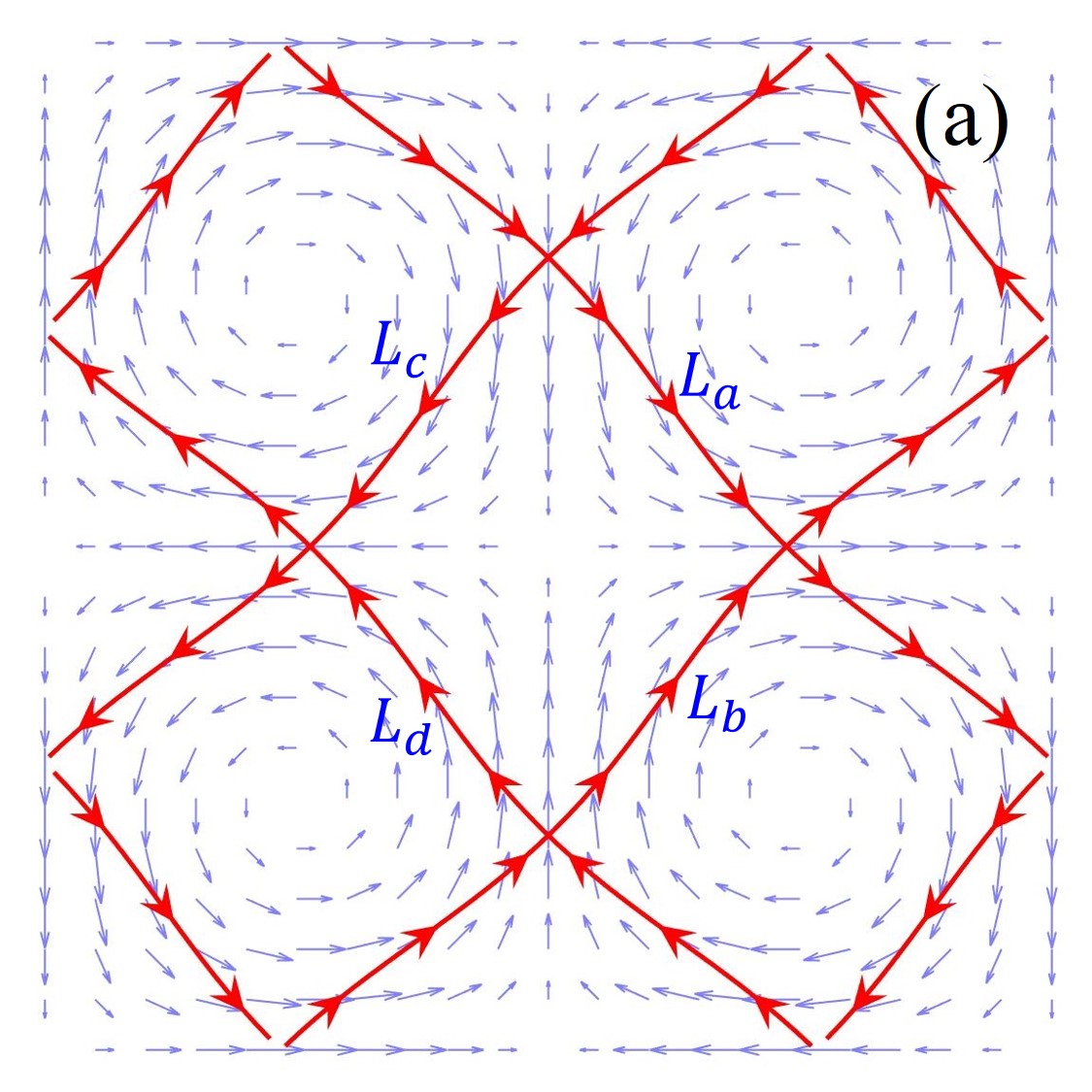}
\includegraphics[width=6cm]{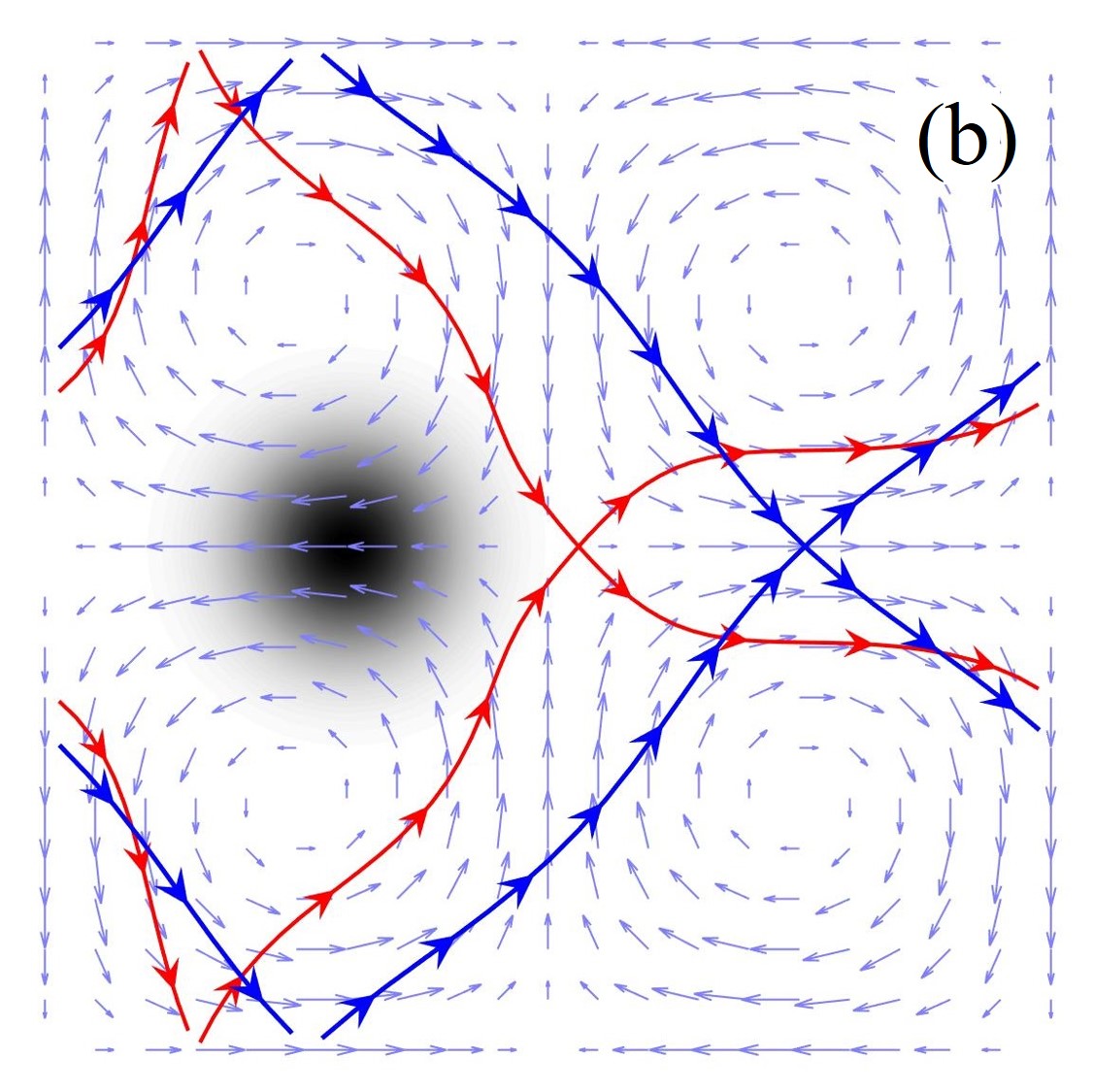}
\includegraphics[width=6cm]{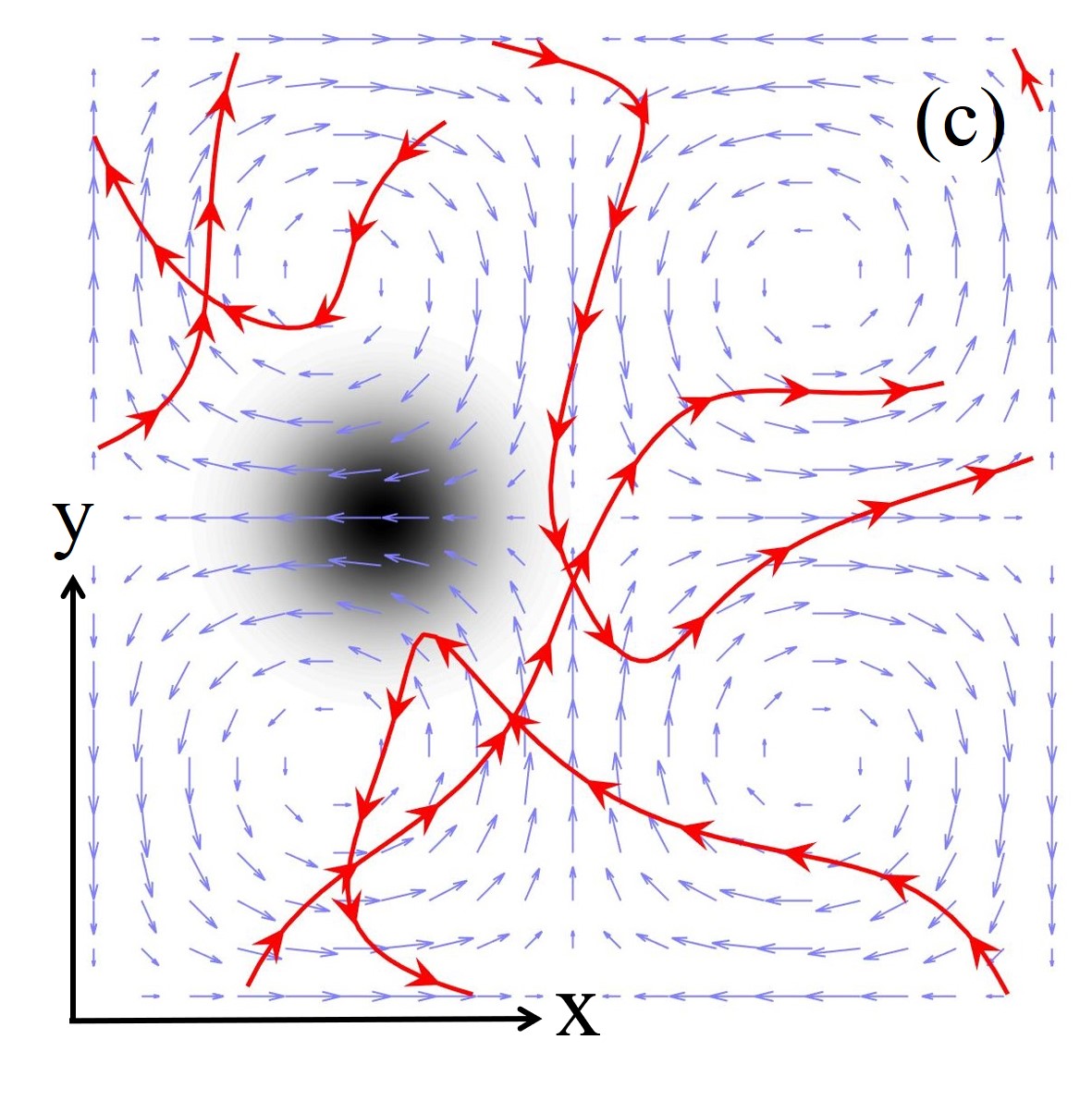}
\includegraphics[width=6cm]{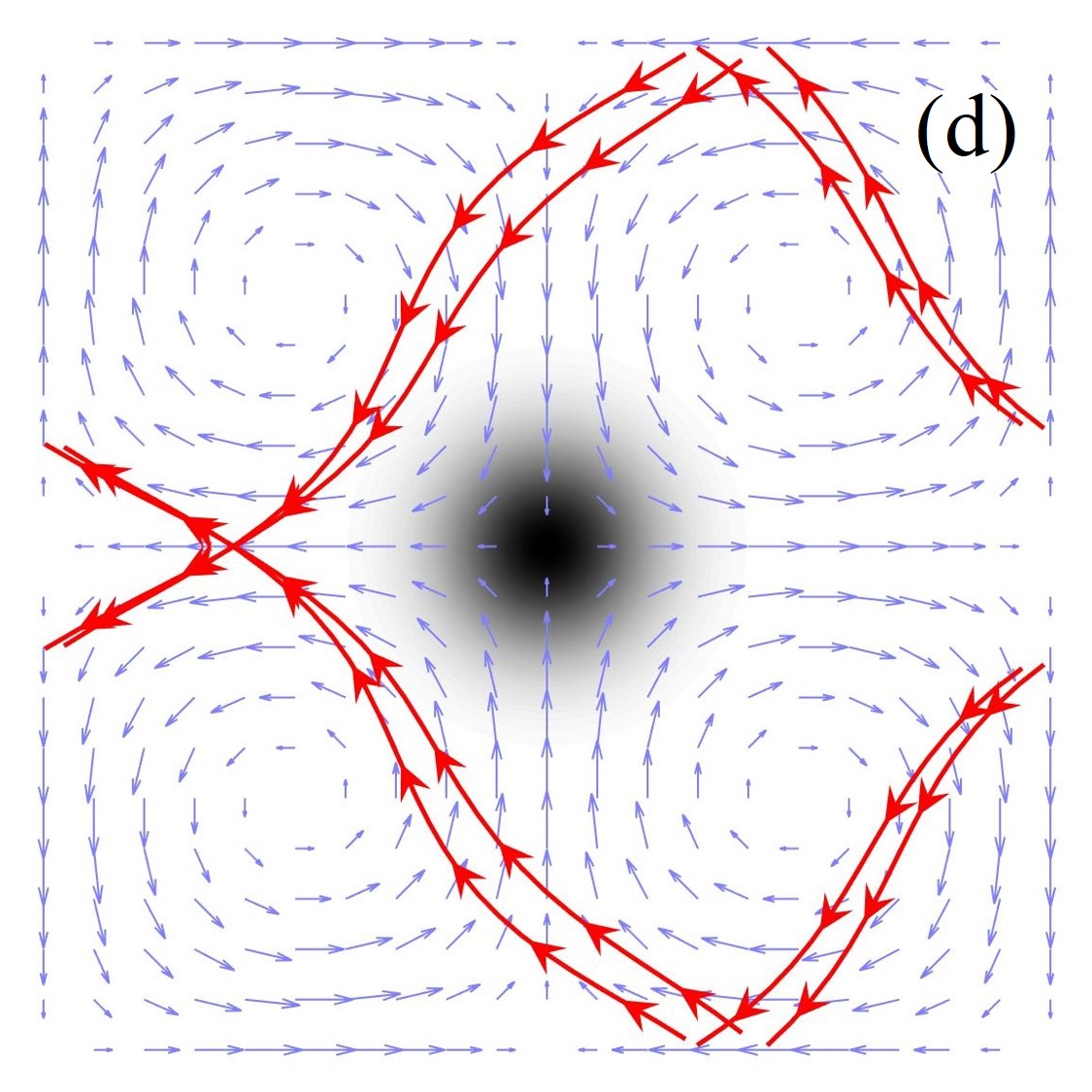}
\caption{
Sample trajectories of a Brownian particle for different cases. (a) $f=A=0.0$, (b) $(f,x_{0})=(-0.2,0.3)$ (red lines) and $(f,x_{0})=(0.0,0.3)$ (blue lines), (c) $(f,x_{0})=(-0.14,0.3)$, (d) $(f,x_{0})=(0.05,0.5)$. The other parameters are same as that in Fig. 2.
}\label{1}
\end{center}
\end{figure}

\indent In order to reveal the transport details, we plot the trajectories of inertial Brownian particles to explain the transport behavior in Fig. 3. When the external force and Gaussian potential are absent, the motion of particles is realized along some preferential channels in the laminar flows\cite{Sarracino}. The rightward channels ($L_a$ and $L_b$) are exactly the same as the leftward channels ($L_c$ and $L_d$), respectively, see Fig. 3(a). However, the presence of Gaussian potential may destroy the generation of some preferential channels. When Gaussian potential are located at $x_{0}=0.3$ [situation $C$ in Fig. 2(a)], the leftward channels ($L_c$ and $L_d$) are destroyed and there are only the rightward channels ($L_a$ and $L_b$), thus the particles move to the right [the blue lines in Fig. 3(b)]. When a constant force $f=-0.2$ is applied to the particles [situation $A$ in Fig. 2(a)], the rightward channels move to the left [the red lines in Fig. 3(b)]. By comparison, the blue lines in Fig. 3(b) are straighter than the red lines, thus the average velocity in situation $C$ is larger than that in situation $A$. When $f=-0.14$ [situation $B$ in Fig. 2(a)], the particles frequently collide with the Gaussian potential, there are no preferential channels, thus the average velocity tends to zero, Fig. 3(c).

\indent In the absence of any potentials, the inertial Brownian particles in cellular flow move with the biased force $f$ except near $f=0.065$ \cite{Sarracino}. In the presence of 2D Gaussian potentials, the particles cannot enter the region influenced by the potentials. Under the combined action of the Gaussian potential and the constant force, the leftward channels shift slightly to the left, while the rightward channels may obviously deviate from the original positions (even be destroyed). When the constant force $f=0.05$, the particles move along the leftward channels, shown in Fig. 3(d). Obviously, the particle can move in a direction opposite to the constant force, i.e., the system exhibits the remarkable phenomenon of ANM, Fig. 2(b).

\indent The above results show apparent advantages over other studies\cite{Sarracino,Cecconi1,Ai2}. To optimize the transport effects, more details will be presented. In the following, two specific cases will be discussed in detail: spontaneous rectification at $f=0.0$ and absolute negative mobility at $x_{0}=0.5$.

\subsection{Spontaneous rectification}

\begin{figure}[htbp]
\begin{center}
\includegraphics[width=8cm]{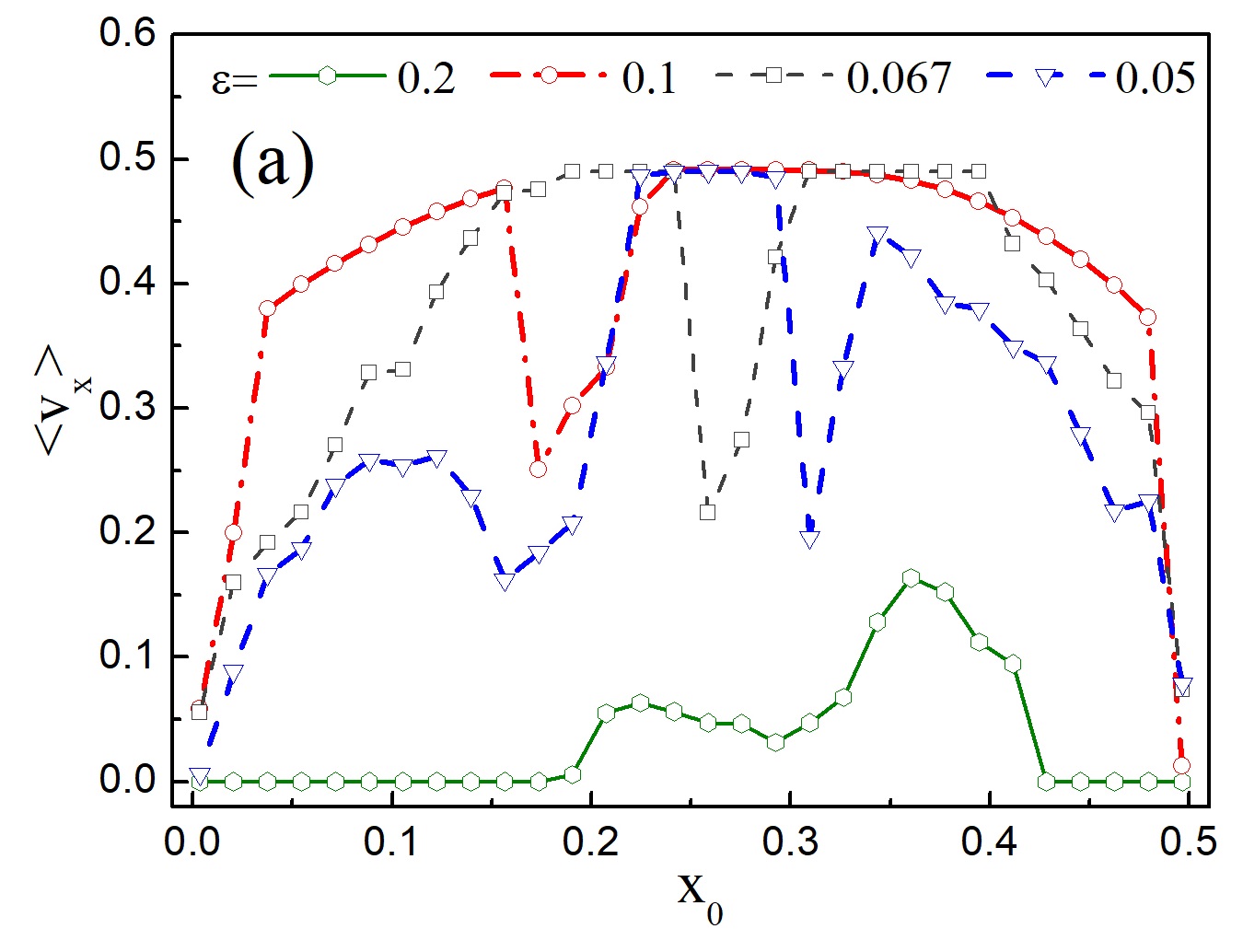}
\includegraphics[width=8cm]{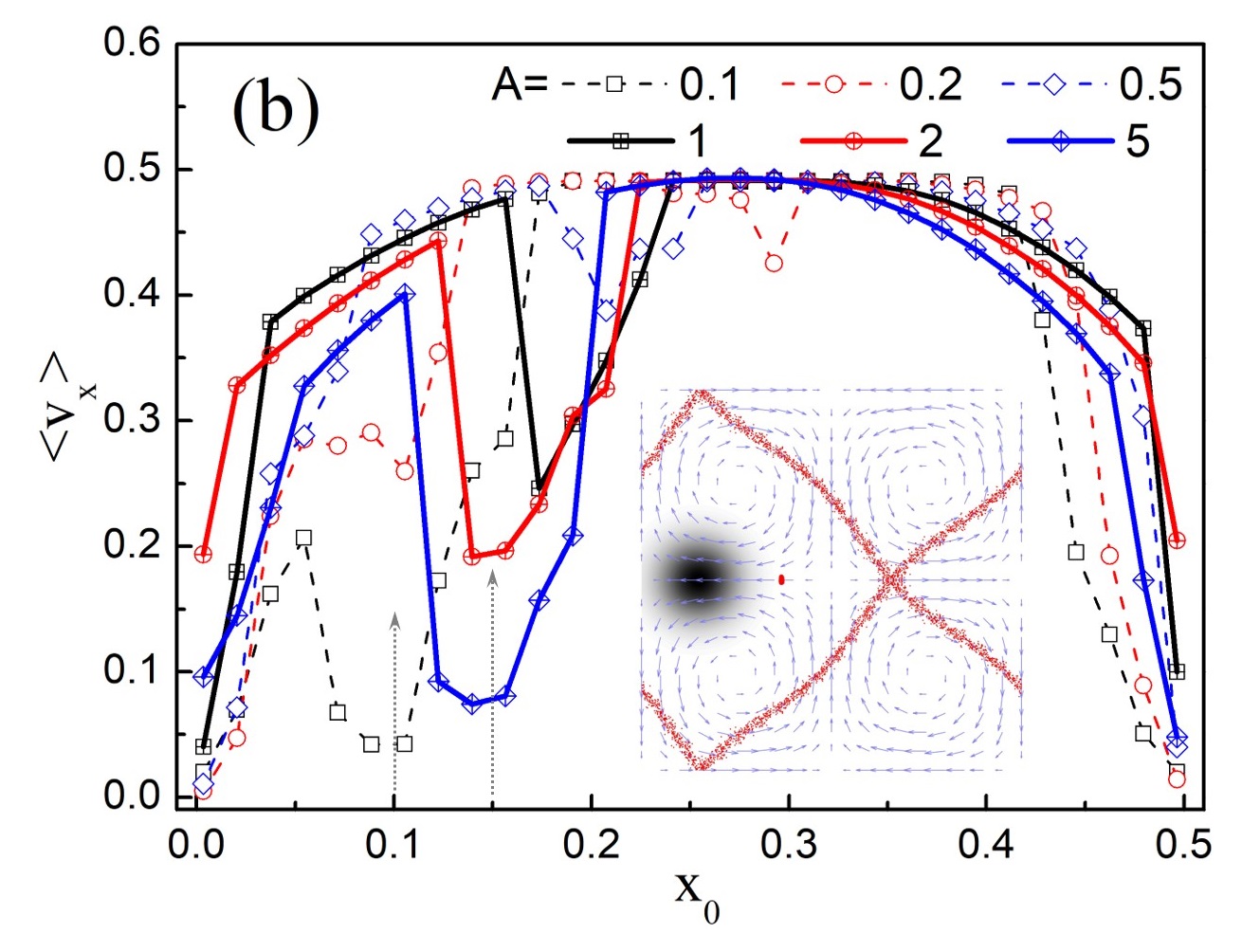}
\caption{
(a) Average velocity $\langle v_{x}\rangle$ as a function of the position $x_{0}$ for different values of $\varepsilon$ at $A=1.0$. (b) Average velocity $\langle v_{x}\rangle$ as a function of the position $x_{0}$ for different values of $A$ at $\varepsilon=0.1$. The other parameters are chosen as $f=0.0$, $D=10^{-5}$ and $\tau=1.0$.
}\label{1}
\end{center}
\end{figure}

\indent Figure 4(a) shows the average velocity $\langle v_{x}\rangle$ as a function of the position $x_{0}$ for different values of $\varepsilon$. When $\varepsilon$ is large, the Gaussian potential will overlap obviously with Gaussian potentials in adjacent cells, thus the particles may be trapped by these potentials. In the case of $\varepsilon=0.2$, the region influenced by the potential is relatively large and the influence of the cellular flow on the rectification becomes weak, thus the average velocity $\langle v_{x}\rangle$ is small. With the decreasing of $\varepsilon$, the region influenced by Gaussian potential decreases. When $\varepsilon$ is very small, the effects of the Gaussian potential can be neglected (not shown here). In the case of $\varepsilon=0.05$, the rectification effect is weak compared to the cases of $\varepsilon=10.0$ and $\varepsilon=15.0$, and the range of $x_{0}$ for obtaining optimal $\langle v_{x}\rangle$ becomes small.

\indent To investigate the effect of potential strength $A$ on the spontaneous rectification, we display the average velocity $\langle v_{x}\rangle$ versus the position $x_0$ for different values of $A$ in Fig. 4(b). It is found that the average velocity is sensitive to the position of Gaussian potential for different $A$. In the case of $x_0=0.1$, the potential will not affect the rightward channels [see Fig. 3(a)], while may have a significant impact on the leftward channels depending on the potential strength $A$. When the potential strength is small (e.g., $A=0.1$), the longitudinal asymmetry is not very obvious and the leftward channels will not be significantly changed, thus the average velocity is small. With the increasing of $A$ $(0.1<A<0.5)$, the Gaussian potentials can cause the system asymmetry and destroy the generation of leftward channels, thus the average velocity increases. However, the average velocity may decrease with further increase of $A$ $(1<A<5)$ due to the existence of trapped particles. The mechanism is the same with the case of $(x_0, A)=(0.15,2)$. In the inset of Fig. 4(b), we plot the distribution of noninteracting Brownian particles at $(x_0, A)=(0.15,2)$. When the potential is at $(x,y)=(0.15,0.5)$, some particles are distributed in the rightward channels. However, the other particles are trapped near $(x,y)=(0.37,0.5)$ (small red region in the inset of Fig. 4(b)) under the combined action of Gaussian potentials and the velocity field.

\begin{figure}[htbp]
\begin{center}
\includegraphics[width=8cm]{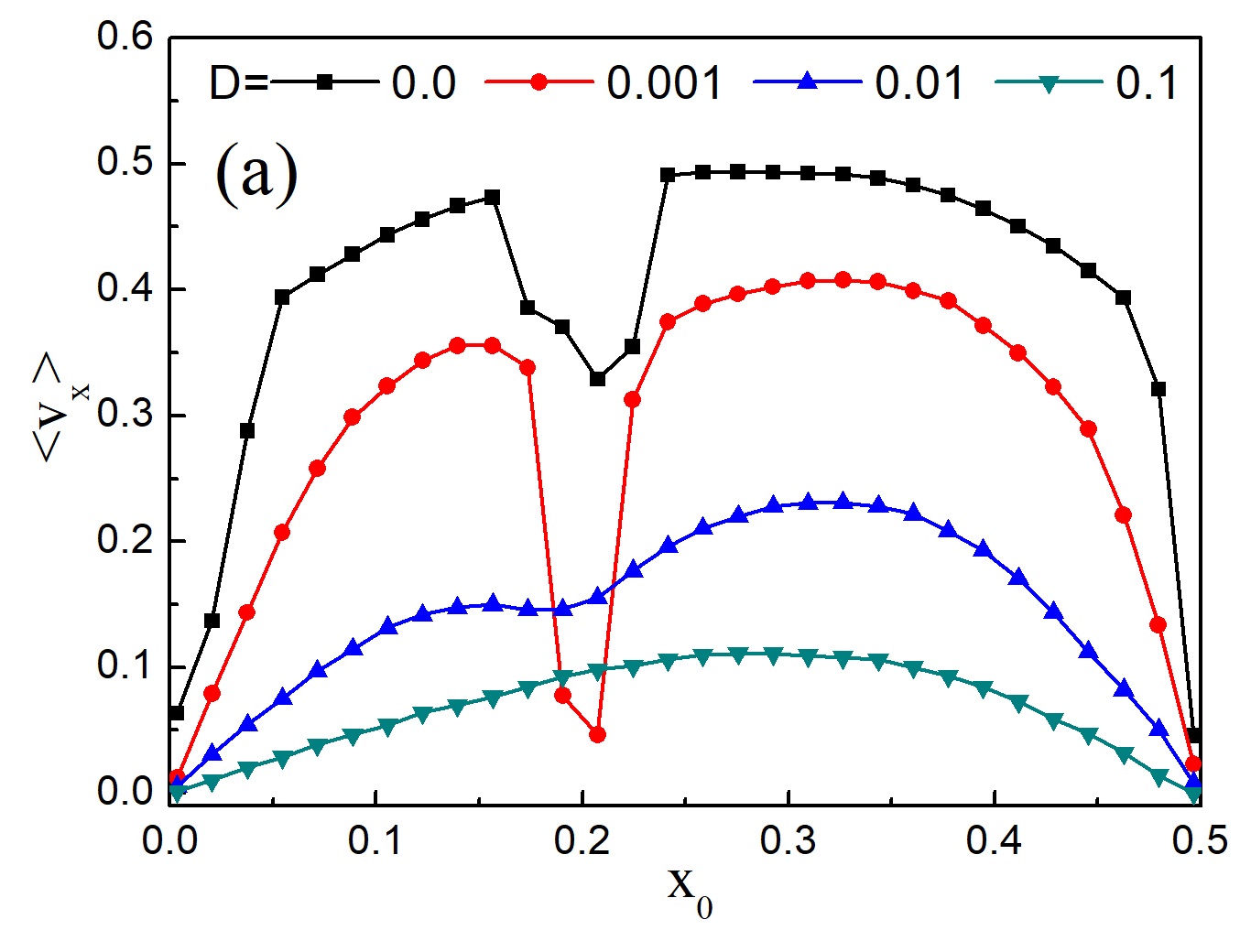}
\includegraphics[width=8cm]{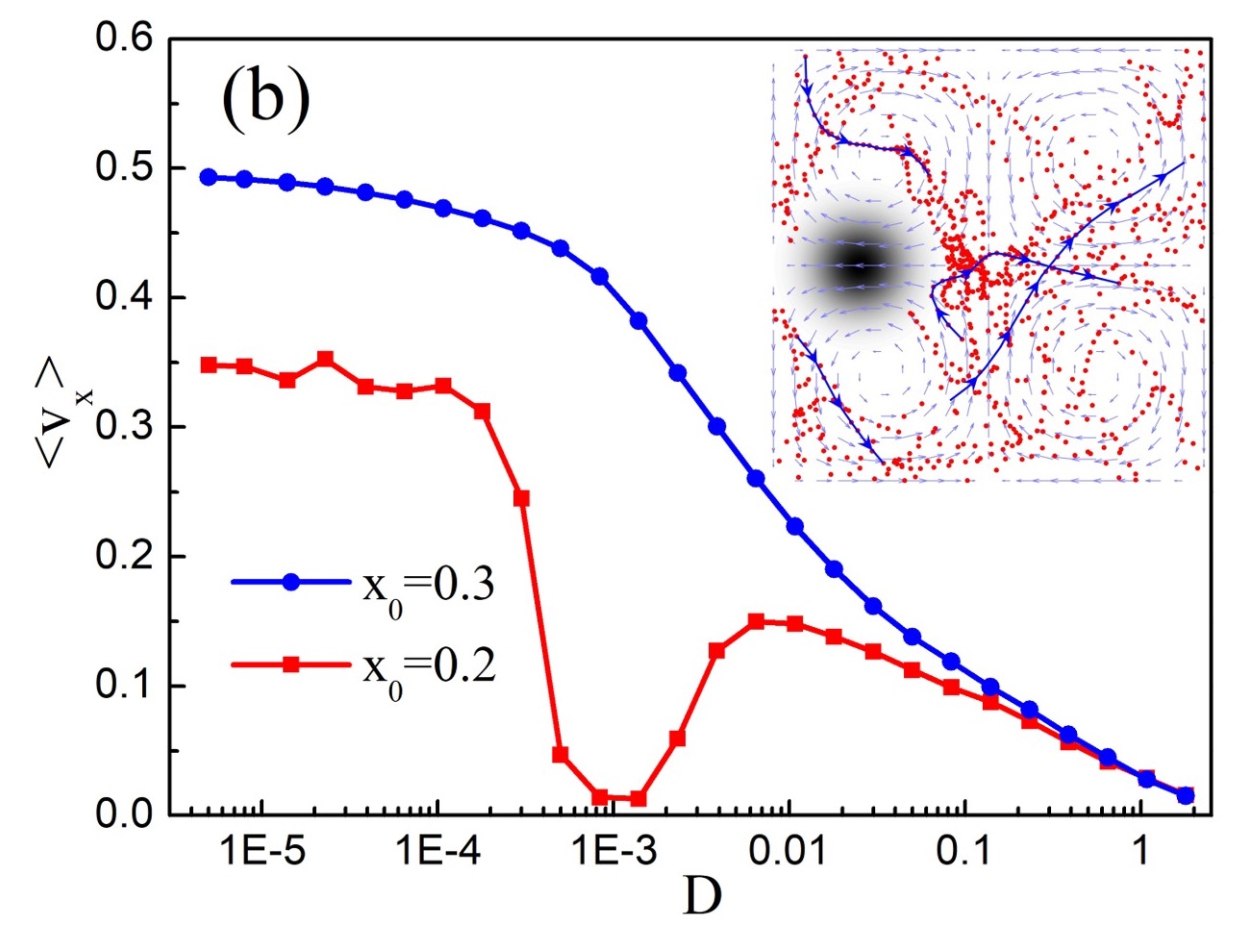}
\caption{
(a) Average velocity $\langle v_{x}\rangle$ as a function of the position $x_{0}$ for different values of $D$. (b) Average velocity $\langle v_{x}\rangle$ as a function of the diffusion coefficient $D$ for different values of $x_{0}$. The parameters are chosen as $f=0.0$, $\varepsilon=0.1$, $A=1.0$ and $\tau=1.0$.
}\label{1}
\end{center}
\end{figure}

\indent The average velocity $\langle v_{x}\rangle$ as a function of the position $x_{0}$ is shown in Fig. 5(a) for different values of $D$. When $D=0$, the system is deterministic, the average velocity $\langle v_{x}\rangle$ shows a nonmonotonic behavior with the increase of $x_{0}$. With the increasing of the diffusion coefficient $D$, the environment noise destroy the stability of the preferential channels, thus the average velocity $\langle v_{x}\rangle$ decreases monotonically, the case of $x_{0}=0.3$ in Fig. 5(b). However, in the case of $x_{0}=0.2$ in Fig. 5(b), the average velocity $\langle v_{x}\rangle$ exhibits very complex behavior. When $D$ is small, most particles move along the rightward channels, while a few are trapped by the Gaussian potentials in laminar flows, thus the average velocity is large. With the increasing of $D$ (e.g., $D=0.001$), the particles move away from the channels, and finally, almost all of them are trapped by the Gaussian potentials, thus the average velocity approaches to zero. When the diffusion coefficient is increased to $D=0.01$, the noise can help the particles to escape from the trapping from the Gaussian potential in laminar flows, thus it may be beneficial to the emergence of SR. In order to explore the physical insights, we plot the inset in Fig. 5(b), which shows the distribution of Brownian particles (see red balls). On the one hand, when the particles enter a small region near $(x,y)=(0.45,0.5)$, they will be trapped by the Gaussian potential. However, the trapping time is very short due to the presence of large noise. On the other hand, although there are no stable preferential channels, the particles are more likely to move to the right (see blue arrows).

\begin{figure}[htbp]
\begin{center}
\includegraphics[width=8cm]{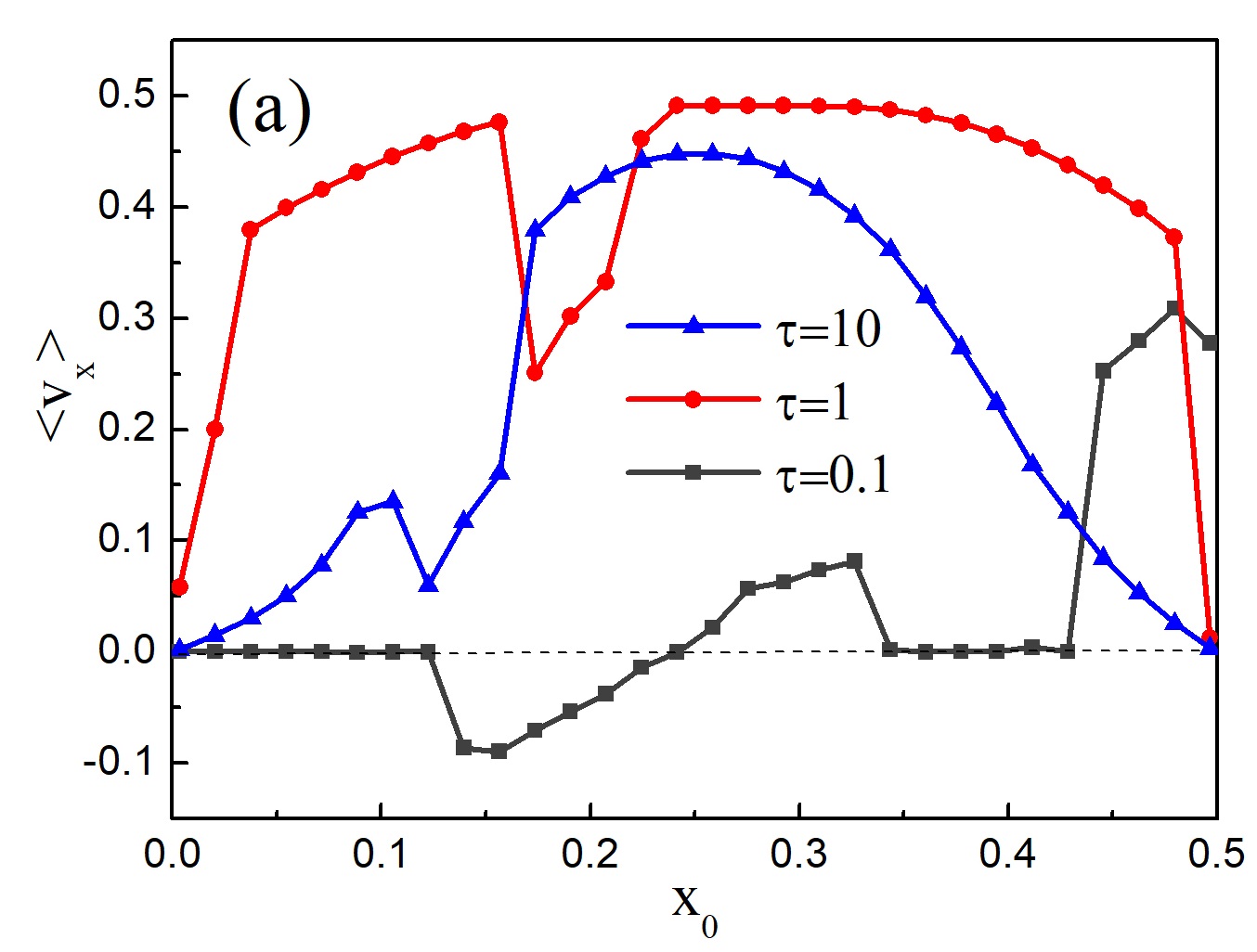}
\includegraphics[width=6cm]{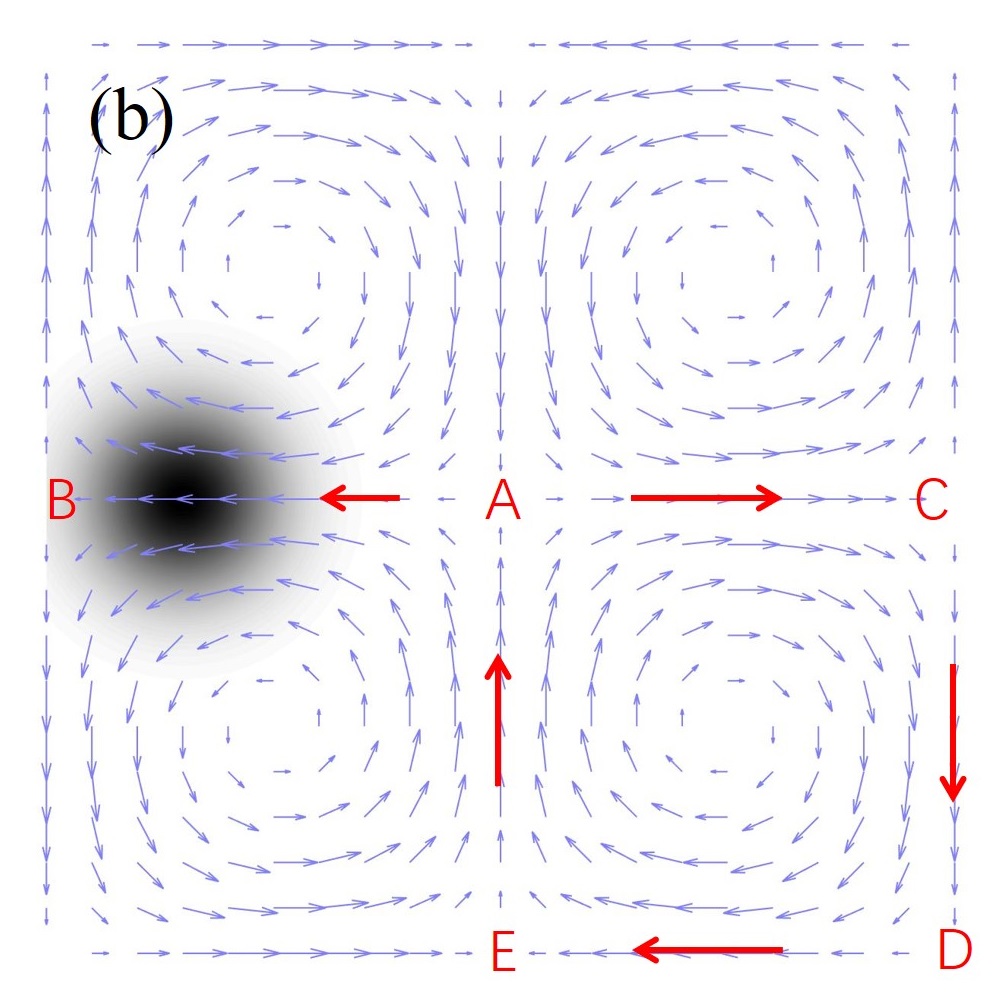}
\caption{
(a) Average velocity $\langle v_{x}\rangle$ as a function of the position $x_{0}$ for different values of $\tau$. (b) Diagrammatic sketch illustrating the rectification of Brownian particles at $\tau=0.1$ and $x_{0}=0.15$, the red arrows indicate the possible direction of the particle movement. The parameters are chosen as $f=0.0$, $\varepsilon=0.1$, $A=1.0$ and $D=10^{-5}$.
}\label{1}
\end{center}
\end{figure}

\indent Figure 6 shows the average velocity $\langle v_{x}\rangle$ as a function of the position $x_{0}$ for different values of $\tau$. It is found that the transport is sensitively dependent to the Stokes time. When the Stokes time is very small (e.g., $\tau=0.1$), the cellular flow induces the particles to move along the velocity field and dominates the transport. Here, we mainly explain the transport in the case of $x_{0}=0.15$. In Fig. 6(b), we plot the diagrammatic sketch at $\tau=0.1$ and $x_{0}=0.15$. We assume the point $A$ is the initial position of the particle. Under the action of the cellular flow, the particles may move along the paths of $A\rightarrow B$ and $A\rightarrow C$. However, the particles can't move from the point $C$ to the point $B$ in adjacent cell due to the double obstructions from the cellular flow and Gaussian Potential. Then the particles may perform the movement along the path of $A\rightarrow C\rightarrow D\rightarrow E\rightarrow A$. Thus the average velocity $\langle v_{x}\rangle$ is negative. When the Stokes time is very large (e.g., $\tau=10.0$), the effects of the cellular flow becomes weak and the average velocity may decrease correspondingly compared to the case of $\tau=1.0$.

\subsection{Absolute negative mobility}

\indent In order to illustrate the effects of Gaussian Potentials on the appearance of ANM, we plot the velocity-force curves and some contour plots of the average velocity versus the system parameters.

\begin{figure}[htbp]
\begin{center}
\includegraphics[width=8cm]{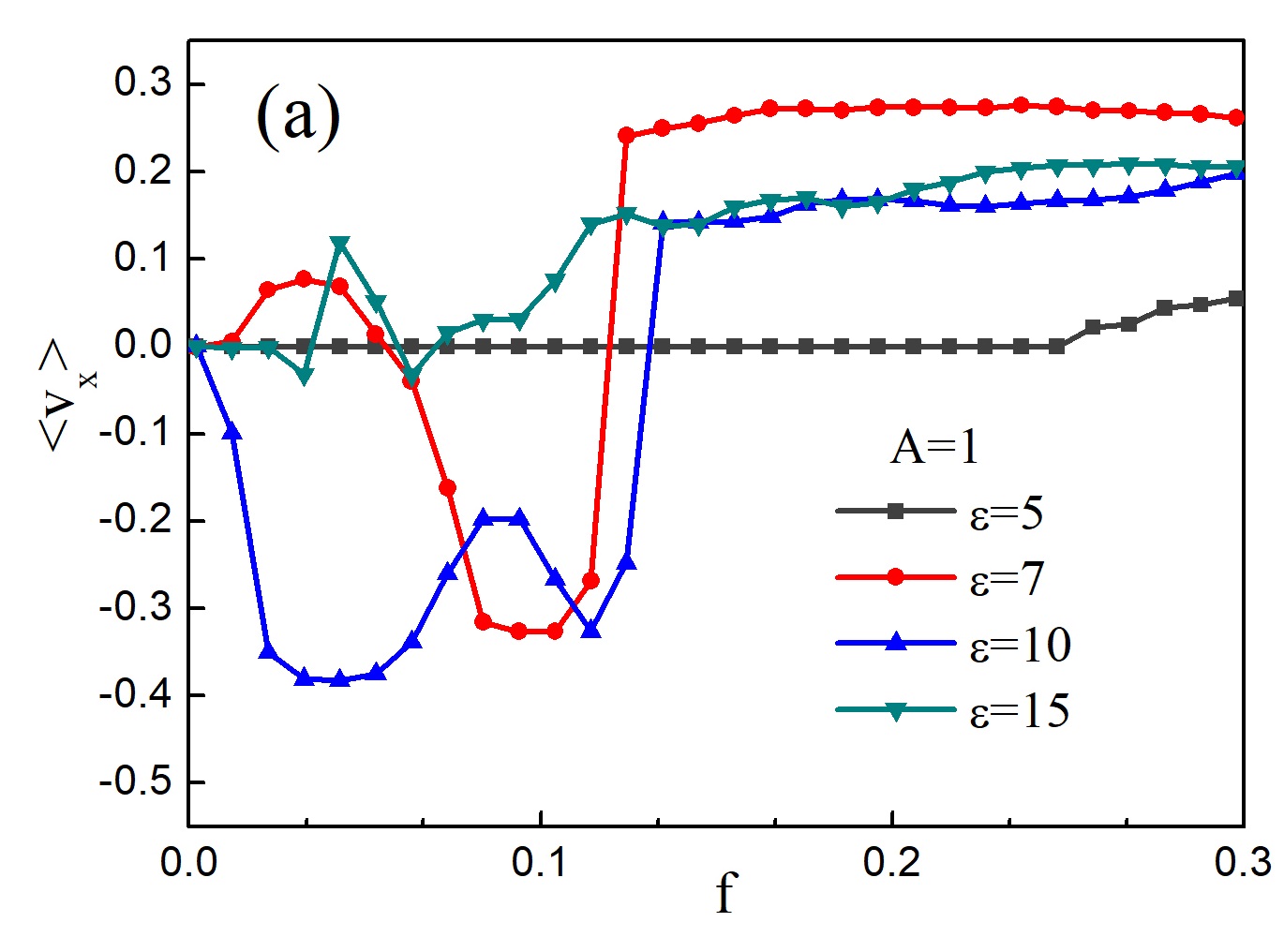}
\includegraphics[width=8cm]{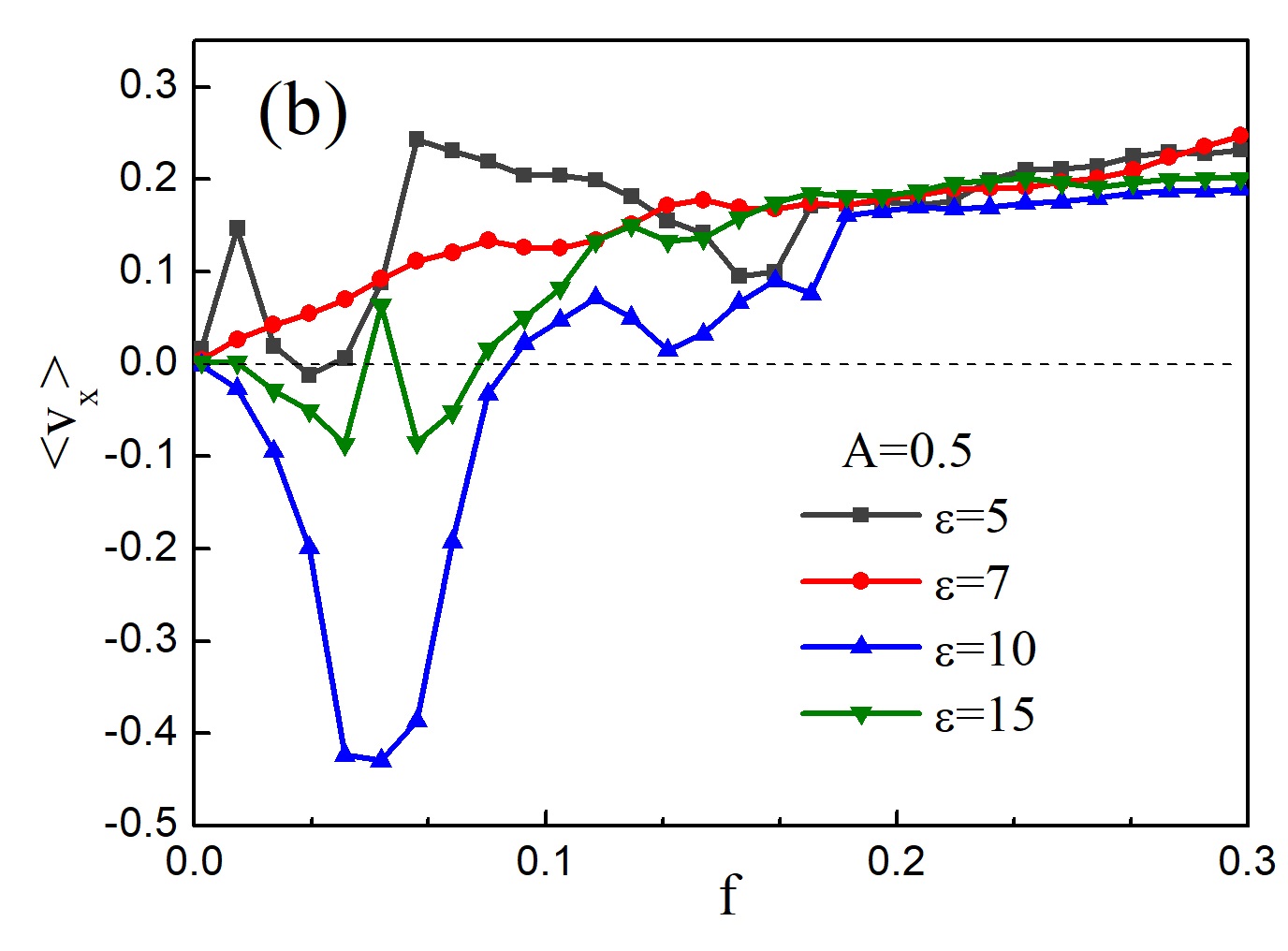}
\includegraphics[width=8cm]{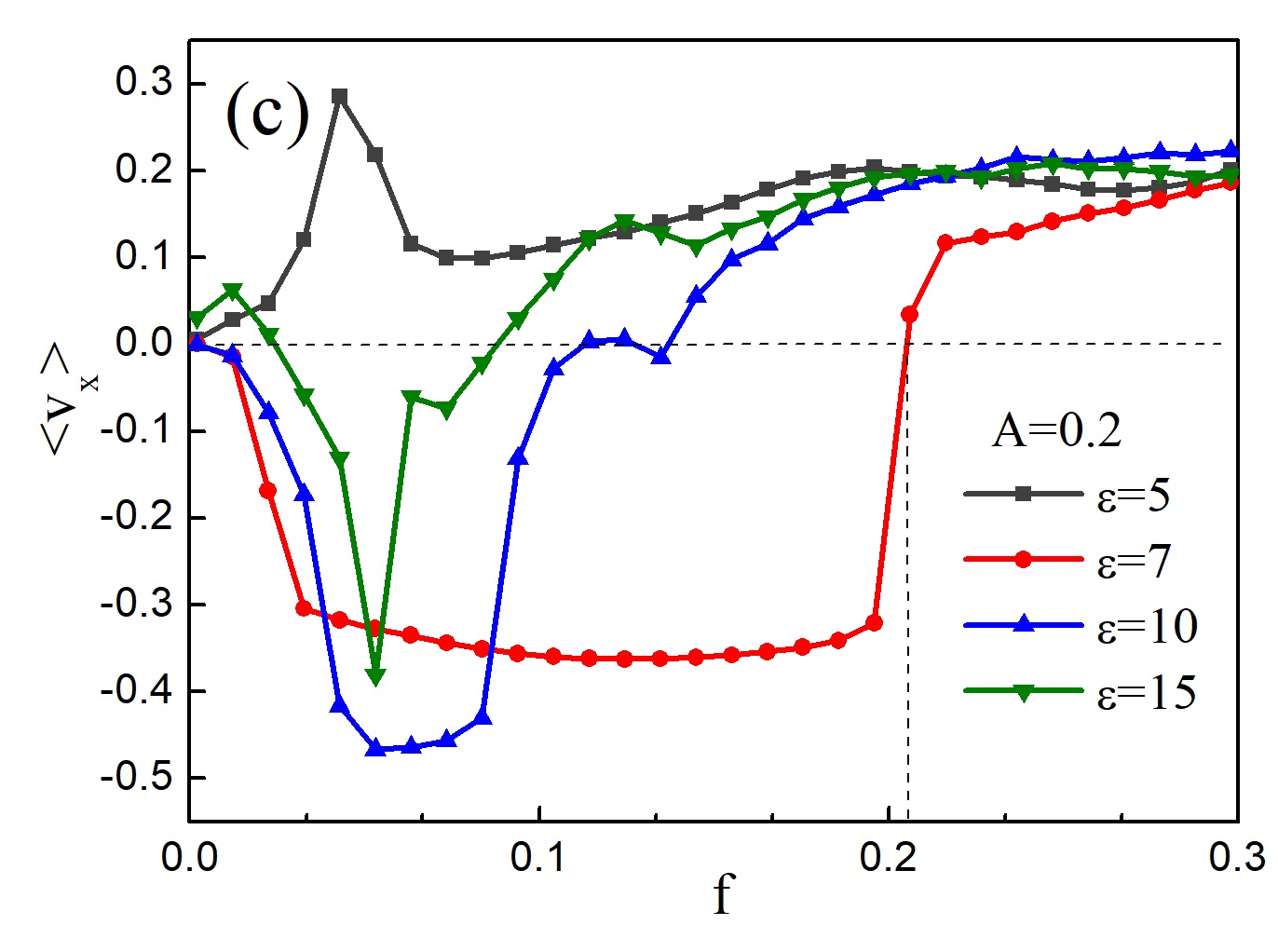}
\includegraphics[width=8cm]{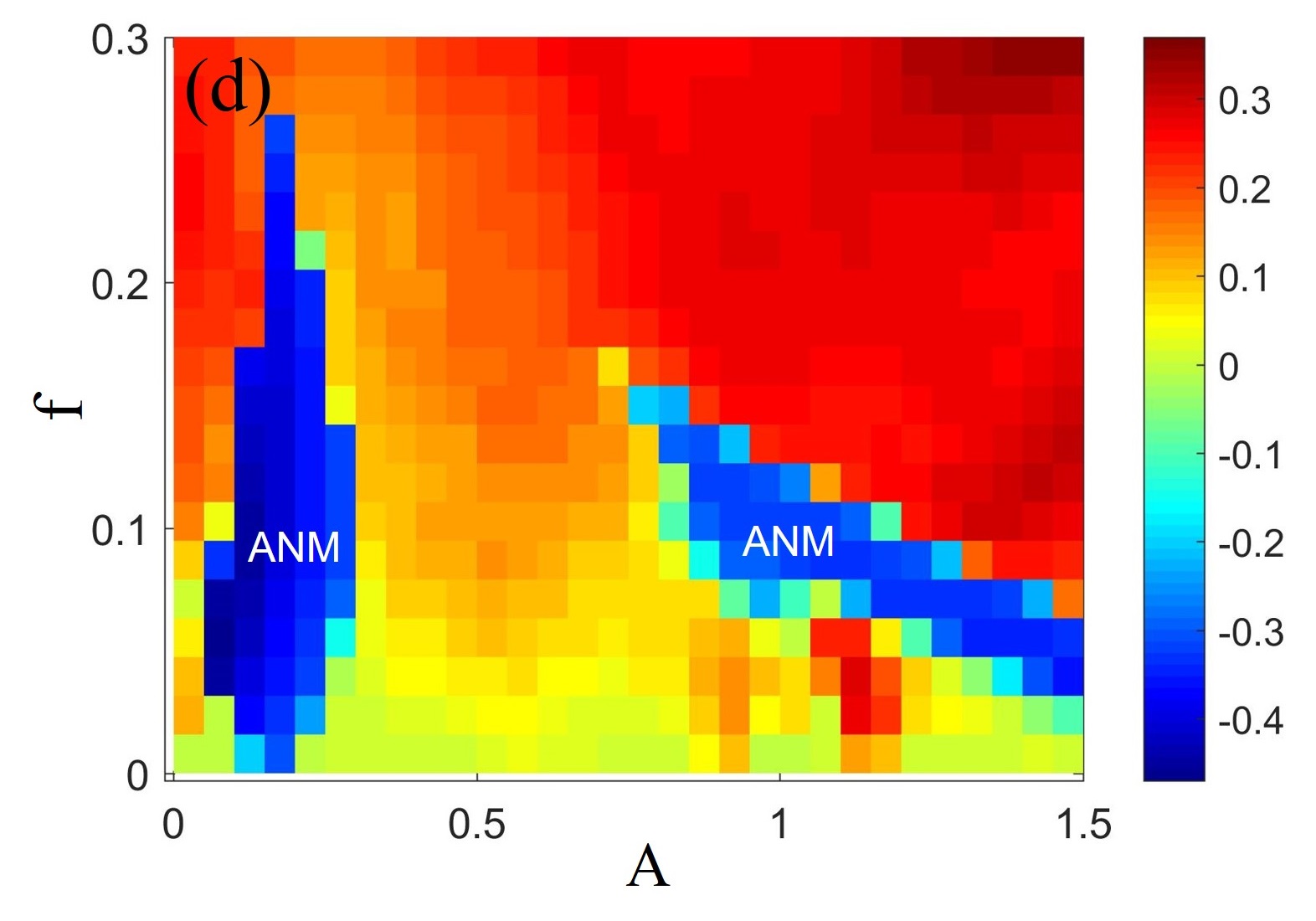}
\caption{
Average velocity $\langle v_{x}\rangle$ as a function of the constant force $f$ for different values of $\varepsilon$ at (a) $A=1.0$, (b) $A=0.5$ and (c) $A=0.2$. (d) Contour plots of the average velocity $\langle v_{x}\rangle$ versus $A$ and $f$ at $\varepsilon=0.143$. The other parameters are chosen as $D=10^{-5}$ and $\tau=1.0$.
}\label{1}
\end{center}
\end{figure}

\indent Figure 7(a) shows the average velocity $\langle v_{x}\rangle$ versus the constant force $f$ for different values of $\varepsilon$ at $A=1.0$. It is found that the velocity-force relation is very complex. When $\varepsilon=0.2$, the particles are easy to be trapped due to the overlap of Gaussian potentials in the case of small constant force $f$. With the decreasing of potential width $\varepsilon$, one can observe the phenomenon of ANM in a certain range of $f$, see the cases of $\varepsilon=0.143$ and $\varepsilon=0.1$. When $\varepsilon$ is small (e.g., $\varepsilon=0.067$), the effect of Gaussian potential becomes weak, thus the regime of $f$ for the appearance of ANM becomes small. Through the comparision of the velocity-force relation in Figs. 7(a-c), it is found that the decreasing of $A$ may be beneficial to the appearance of ANM under appropriate conditions. In the case of $A=0.2$, the amplitude of ANM is greatly improved at $f=0.05$ and $\varepsilon=0.1$, and the regime of $f$ for the appearance of ANM is further increased compared to that in Fig. 2(b) (see Fig. 7(c)).

\indent In order to investigate the dependence of $\langle v_{x}\rangle$ on $A$ and $f$ in details, we plot the contour plots $\langle v_{x}\rangle$ versus the strength of the potential $A$ and the constant force $f$ at $\varepsilon=0.143$ in Fig. 7(d). There exist two regions for the appearance of ANM. One can observe a remarkable phenomenon of ANM in the range of $0.05<A<0.3$ compared to the case of $A>0.75$. In the following discussions, we will choose the parameters $0.05<A<0.3$ and $\varepsilon=0.143$.

\begin{figure}[htbp]
\begin{center}
\includegraphics[width=8cm]{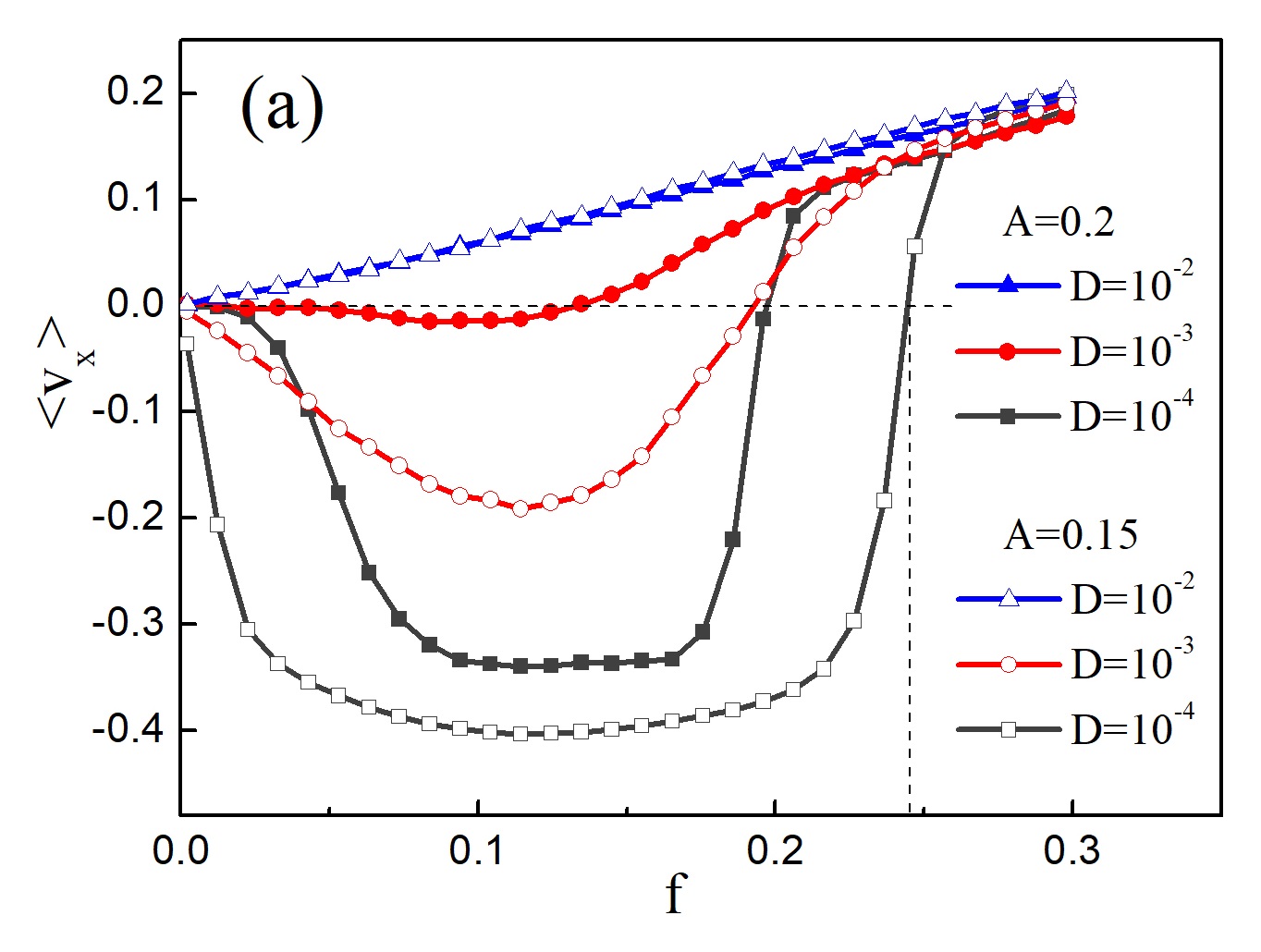}
\includegraphics[width=8cm]{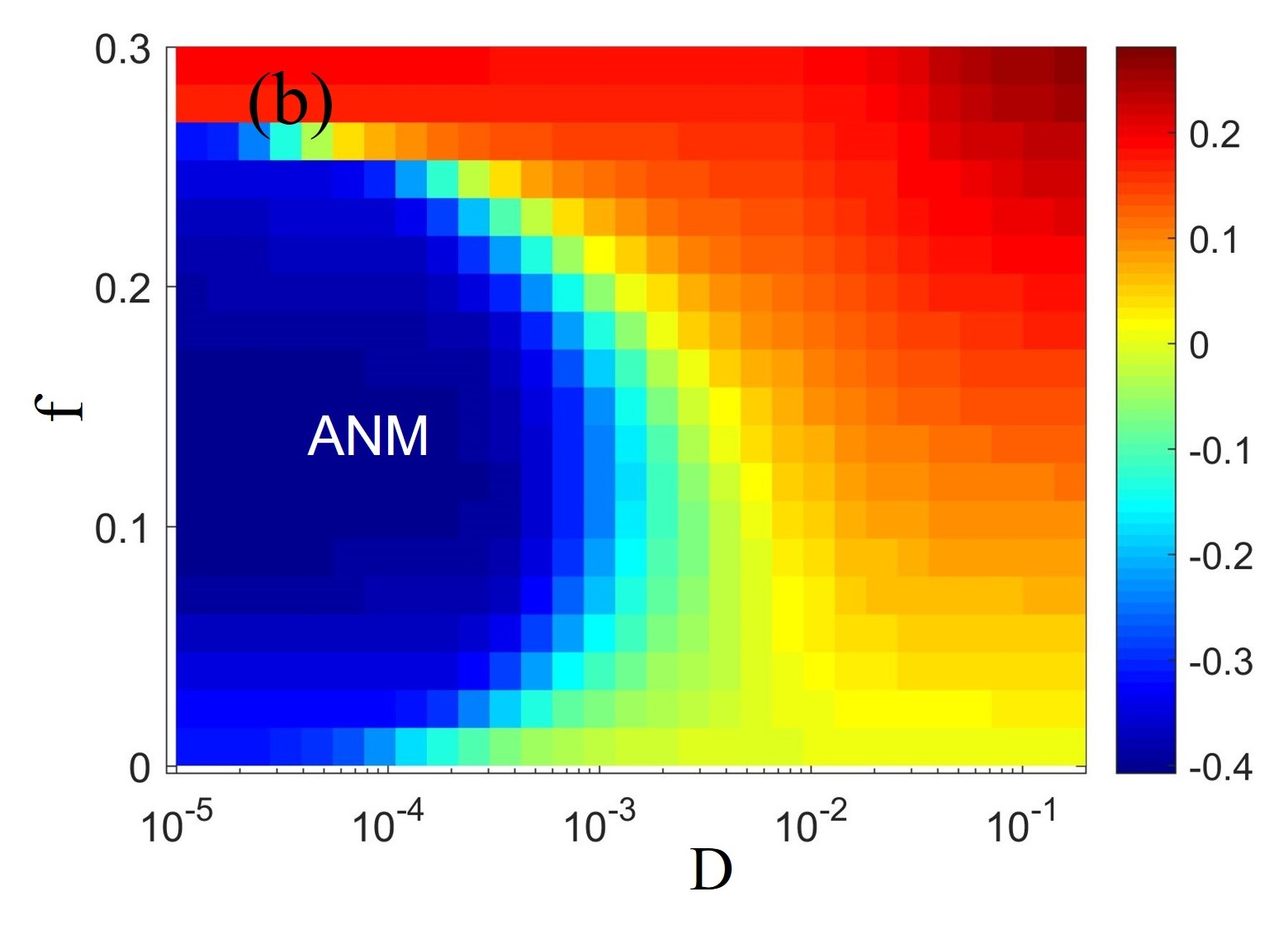}
\caption{
(a) Average velocity $\langle v_{x}\rangle$ as a function of the constant force $f$ for different values of $(A,D)$. (b) Contour plots of the average velocity $\langle v_{x}\rangle$ versus $D$ and $f$ at $A=0.15$. The other parameters are chosen as $\varepsilon=0.143$ and $\tau=1.0$.
}\label{1}
\end{center}
\end{figure}

\indent The random fluctuations may play a key role for the appearance of ANM\cite{Machura}, while suppress this phenomenon in steady laminar flows\cite{Sarracino,Cecconi1,Cecconi2,Ai2}. The dependence of $\langle v_{x}\rangle$ on the constant force $f$ for different $D$ is presented in Fig. 8(a). When the diffusion coefficient is large enough (e.g., $D=0.01$), the effect
of the velocity field is averaged out, thus the average velocity $\langle v_{x}\rangle$ increases monotonically with the constant force $f$. When the diffusion coefficient is small (e.g., $D=10^{-4}$ and $10^{-3}$), the phenomenon of ANM can be observed. By contrast, a very obvious phenomenon of ANM can be observed by choosing $A=0.15$. To show the dependence of $\langle v_{x}\rangle$ on $D$ and $f$ in more details, we plot the contour plots $\langle v_{x}\rangle$ versus $D$ and $f$ at $A=0.15$ in Fig. 8(b). It is found that ANM can occur in the case of $D<0.004$ and the range of $f$ for obtaining ANM increases with the decreasing of $D$. These results indicate that the random fluctuations may reduce the system nonequilibrium nature from the velocity field and a very obvious phenomenon of ANM appears for small thermal fluctuations.

\begin{figure}[htbp]
\begin{center}
\includegraphics[width=8cm]{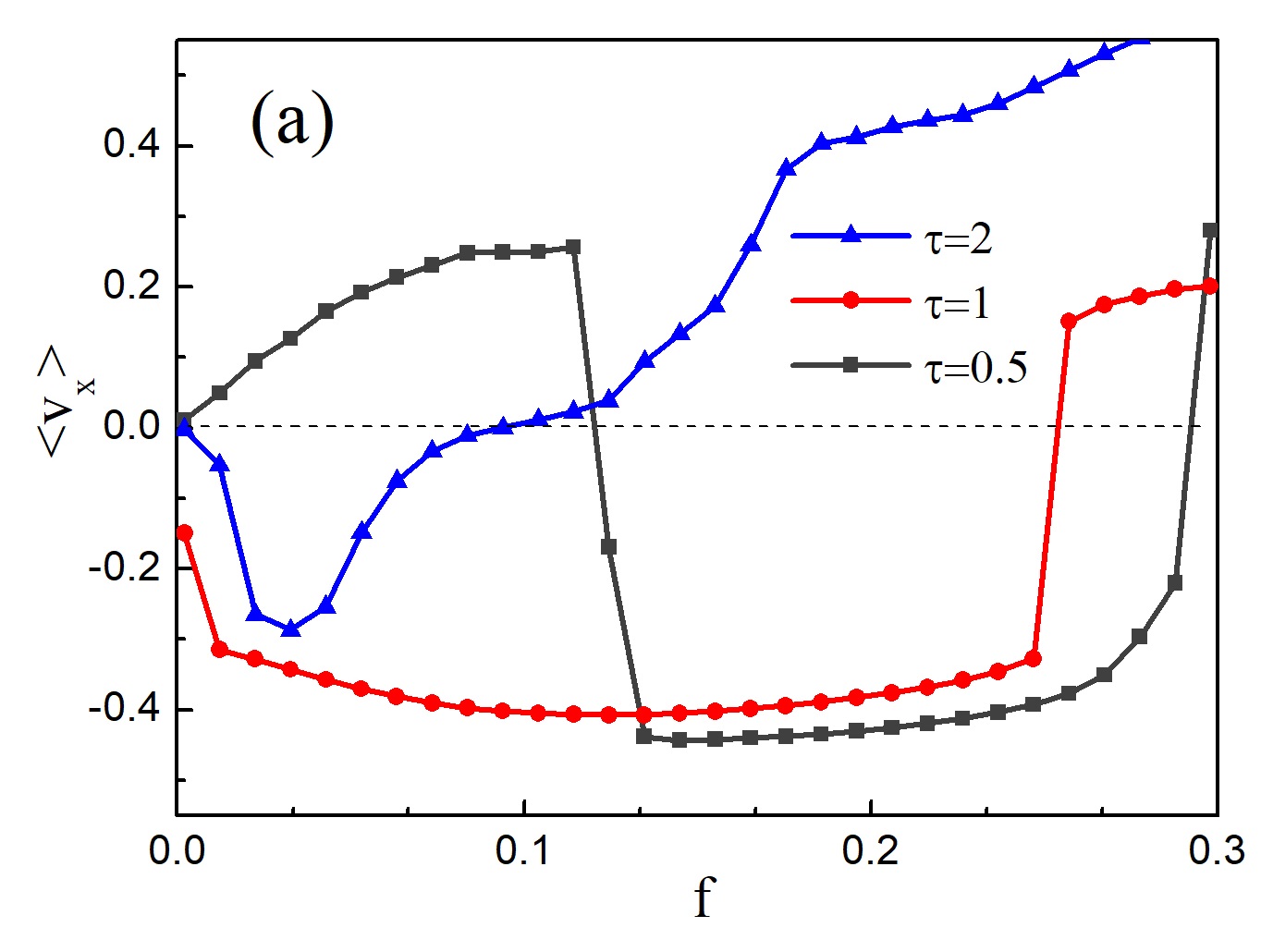}
\includegraphics[width=8cm]{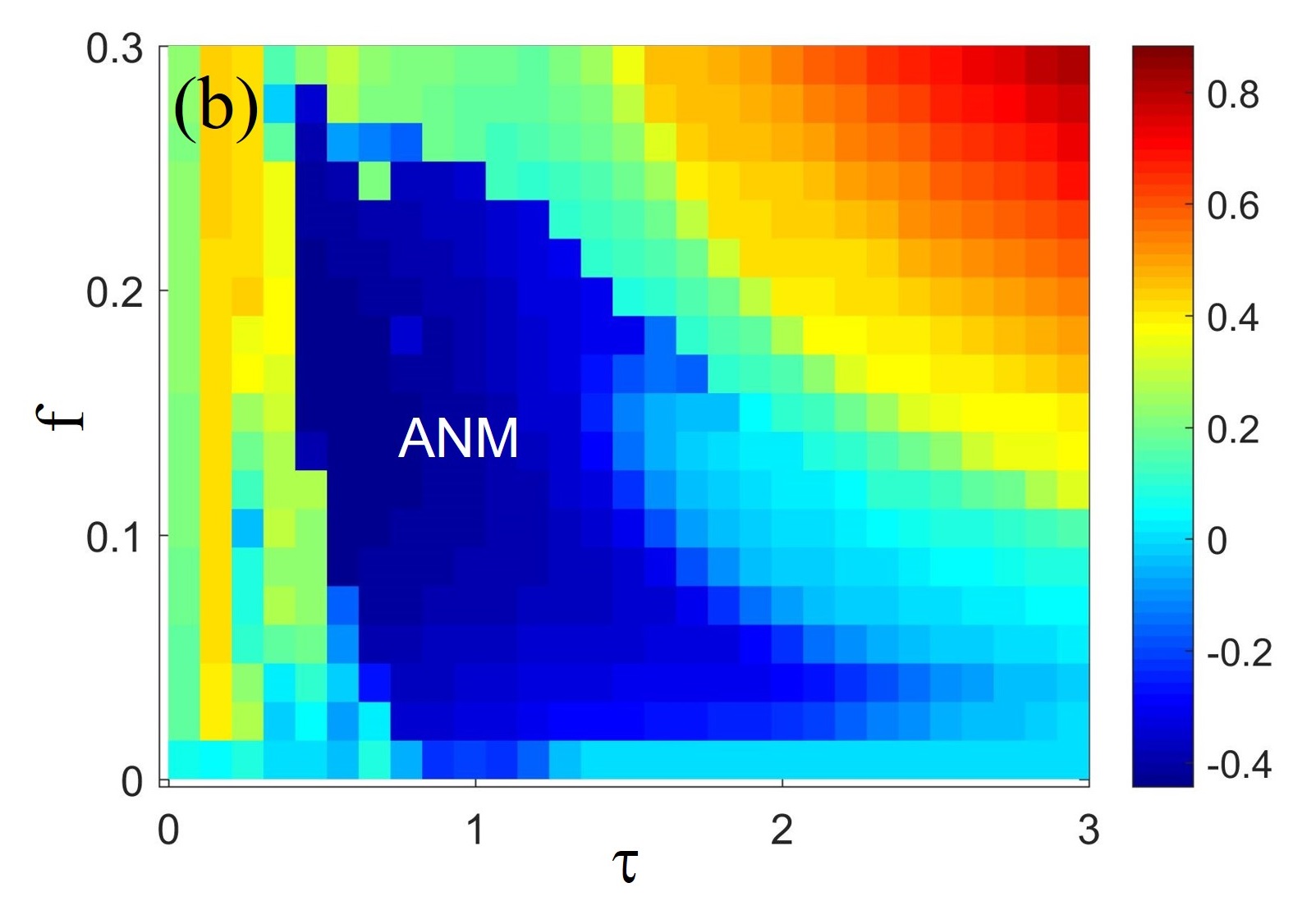}
\caption{
(a) Average velocity $\langle v_{x}\rangle$ as a function of the constant force $f$ for different values of $\tau$. (b) Contour plots of the average velocity $\langle v_{x}\rangle$ versus $\tau$ and $f$. The other parameters are chosen as $A=0.15$, $\varepsilon=0.143$ and $D=10^{-5}$.
}\label{1}
\end{center}
\end{figure}

\indent In Fig. 9(a), we investigate the average velocity $\langle v_{x}\rangle$ versus the constant force $f$ for different values of $\tau$. It is found that a proper value of $\tau$ can promote the appearance of ANM, see the case of $\tau=1.0$. Moreover, the regime of $f$ for the appearance of ANM is increased by a factor of 7 compared to the Ref.\cite{Ai2}. However, when the Stokes time is small or large, the regimes of $f$ for the appearance of ANM are both small, see the cases of $\tau=0.5$ and $\tau=2.0$. The detailed contour plots $\langle v_{x}\rangle$ versus $\tau$ and $f$ are plotted in Fig. 9(b). It is found that ANM may occur in a range of $0.4<\tau<3.0$. On increasing $\tau$ from 1.0 to 3.0, the regime of $f$ for the appearance of ANM reduces and finally disappears.

\begin{figure}[htbp]
\begin{center}
\includegraphics[width=8cm]{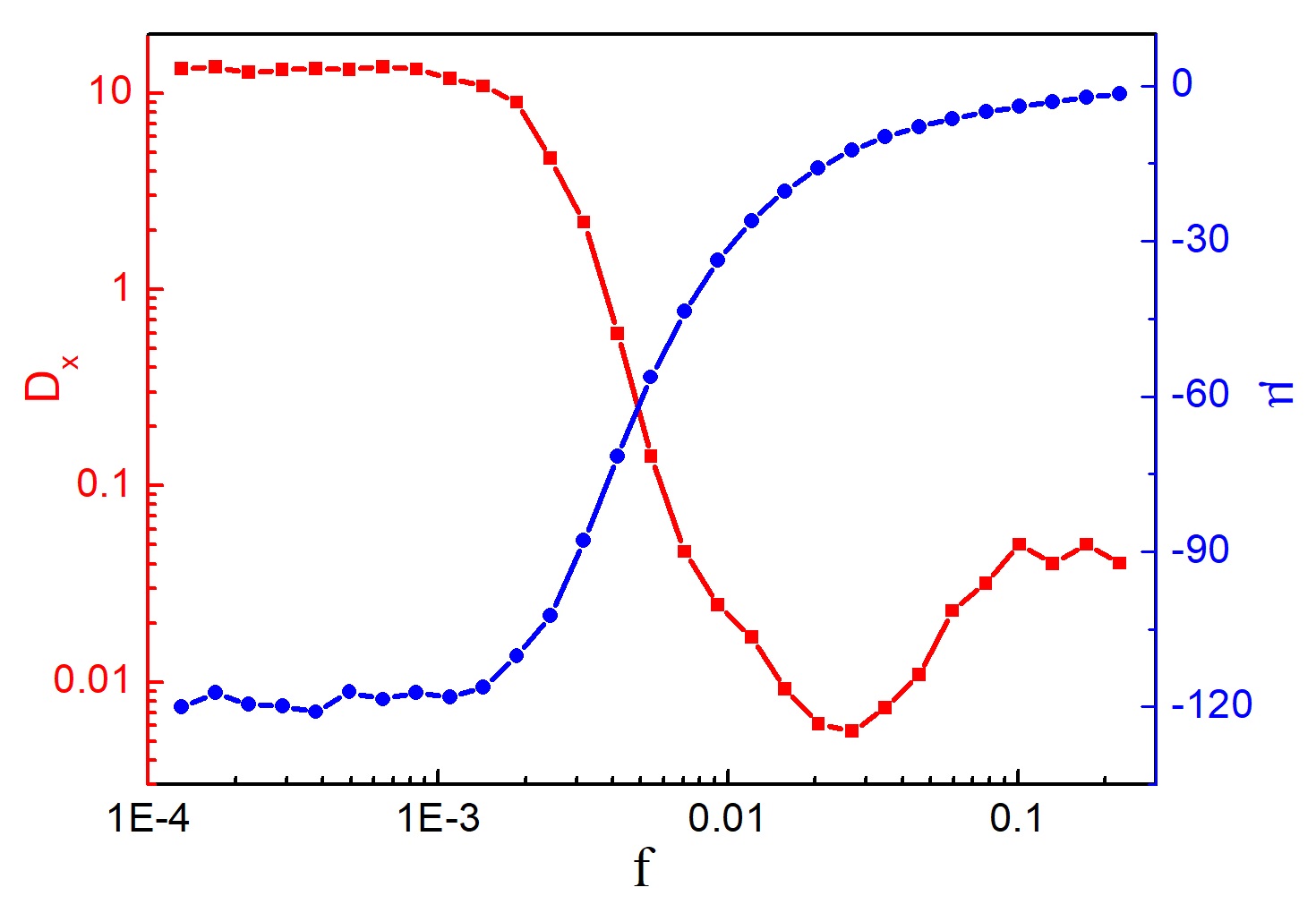}
\caption{
Effective diffusion coefficient $D_x$ and mobility $\mu$ as a function of the constant force $f$ for the case of $\tau=1.0$ in Fig. 9(a).
}\label{1}
\end{center}
\end{figure}

\indent To obtain the necessary condition for the appearance of ANM in the present system, we display the effective diffusion coefficient $D_{x}$ and mobility $\mu$ versus $f$ ($f<0.25$). When $f$ is small, both $D_{x}$ and $\mu$ are nearly independent of the constant force $f$. With the increasing of $f$, the amplitude of negative mobility decreases monotonically. However, $D_x$ first decreases to its minimal value, then increases to an extremum value near $f = 0.25$. Because the preferential channels in the present system change near the critical bias, thus the direction of particle mobility change and the effective diffusion coefficient take its maximum value. When $f$ is large enough, the effects of the Gaussian potentials and the velocity field are negligible, $\mu\rightarrow 1$ and $D_{x}\rightarrow D$\cite{Sarracino}. Compared with previous results\cite{Sarracino,Ai2}, the regime of $f$ for the appearance of ANM increases significantly and is accompanied by low $D_{x}$. Therefore, it is possible to obtain obvious ANM by reducing the effective diffusion coefficient.

\section {Concluding remarks}
\indent In conclusion, we numerically investigated the transport of inertial Brownian particles induced by Gaussian potentials in steady laminar flows. We found that the transport is sensitively dependent on the Gaussian potential and the external constant force. In the absence of any external driving forces, the Gaussian potential produces the asymmetry of the system and induces the spontaneous rectification of particles. The simulation results show that the rectification can be strongly enhanced by applying a proper Gaussian potential and the average velocity can approach 0.5. Generally speaking, the average velocity decreases monotonically with the increasing of the diffusion coefficient. However, when the Gaussian potential is located at $x_{0}=0.2$, the particles may be trapped by the potential or escape from it under different random fluctuations, thus the average velocity exhibits a nonmonotonic behavior. When the Stokes time is small ($\tau=0.1$), the transport may reverse its direction under the combined action of the Gaussian Potential and the cellular flow. In addition, when the potential is located at the center of the cellular flow, the system exhibits the phenomenon of ANM under proper parameter conditions. By adjusting the strength of the potential ($0.05<A<0.3$ and $A>0.75$), absolute negative mobility occurs and the regime of $f$ for the appearance of ANM is very large, e.g., $f<0.2$ at $A=0.2$ and $\varepsilon=0.143$. When $A=0.15$ and $\varepsilon=0.143$, the regime of $f$ for the appearance of ANM can be further expanded for small thermal fluctuations and increased by a factor of 7 compared to the previous study\cite{Ai2}. Finally, we discuss the possible experimental implementation. The steady laminar flows can be realized in a setup with rotating cylinders \cite{Solomon}. The inertial Brownian particles interact with the repulsive Gaussian potential obtained by using Gaussian laser beam \cite{Marchant}.

\indent This work was supported in part by the National Natural Science Foundation of China (Grants No. 11747109 and 51576105), the Natural Science Foundation of Jiangxi Province (Grant No. 20181BAB211011), the Science Foundation of Jiangxi Provincial
Department of Education (Grant No. Gjj161057 and No. Gjj14718).

\end{document}